\documentclass[final,onefignum,onetabnum]{siamart190516}

\usepackage{lipsum}
\usepackage{amsfonts}
\usepackage{graphicx}
\usepackage{epstopdf}
\usepackage{algorithmic}
\ifpdf
  \DeclareGraphicsExtensions{.eps,.pdf,.png,.jpg}
\else
  \DeclareGraphicsExtensions{.eps}
\fi

\usepackage{url}

\usepackage{xspace}
\usepackage{numprint}
\usepackage{amsopn}
\usepackage{float}

\usepackage{multirow}
\usepackage{ctable}
\usepackage{rotating}
\usepackage{lscape}
\usepackage{array}
\usepackage{booktabs}
\usepackage{supertabular}
\usepackage{tabularx}

\usepackage{enumitem}
\setlist[enumerate]{leftmargin=.5in}
\setlist[itemize]{leftmargin=.5in}

\usepackage[]{xcolor}

\newsiamremark{remark}{Remark}
\newsiamremark{hypothesis}{Hypothesis}
\crefname{hypothesis}{Hypothesis}{Hypotheses}
\newsiamthm{claim}{Claim}

\headers{Industry-Relevant Implicit Large-Eddy Simulation}{G. Mengaldo, et al.}

\title{Industry-Relevant Implicit Large-Eddy Simulation of a High-Performance Road Car via Spectral/$hp$ Element Methods 
\thanks{Submitted to the editors DATE.
\funding{EPSRC under the Platform Grant PRISM: Underpinning Technologies for Finite Element Simulation (EP/L000407/1); EU Horizon 2020 project ExaFLOW (grant 671571). EPSRC and Airbus under an Industrial CASE studentship.
}
}
}

\author{
Gianmarco Mengaldo\thanks{National University of Singapore (NUS), Singapore (\email{mpegim@nus.edu.sg})} 
\and 
David Moxey\thanks{University of Exeter, UK (\email{d.moxey@exeter.ac.uk})}
\and
Michael Turner\thanks{Imperial College London, UK (\email{mturner4350@gmail.com}, \email{j.peiro@imperial.ac.uk}, \email{s.sherwin@imperial.ac.uk})}
\and
Rodrigo C. Moura\thanks{Instituto Tecnol\'ogico de Aeron\'autica, Brazil (\email{moura@ita.br})}
\and
Ayad Jassim\thanks{Hewlett Packard Enterprise, UK (\email{ayad.jassim@hpe.com})}
\and
Mark Taylor\thanks{London Computational Solutions, UK (\email{mark@lcs-fast.com})}
\and
Joaquim Peir\'o\footnotemark[4]
\and 
Spencer J. Sherwin\footnotemark[4]
}

\newcommand{\ignore}[1]{}

\newcommand{\nekpp}{\texttt{Nektar++}\xspace}
\newcommand{\nm}{\texttt{NekMesh}\xspace}

\newcommand{\cf}{\texttt{CADfix}\xspace}

\newcommand{\ccm}{\texttt{Star-CCM+}\xspace}

\begin{document}

\maketitle

\begin{abstract}
We present a successful deployment of high-fidelity Large-Eddy Simulation (LES) technologies based on spectral/$hp$ element methods to industrial flow problems, which are characterized by high Reynolds numbers and complex geometries. In particular, we describe the numerical methods, software development and steps that were required to perform the implicit LES of a real automotive car, namely the Elemental Rp1 model. To the best of the authors' knowledge, this simulation represents the first high-order accurate transient LES of an entire real car geometry. Moreover, this constitutes a key milestone towards considerably expanding the computational design envelope currently allowed in industry, where steady-state modelling remains the standard. To this end, a number of novel developments had to be made in order to overcome obstacles in mesh generation and solver technology to achieve this simulation, which we detail in this paper. The main objective is to present to the industrial and applied mathematics community, a viable pathway to translate academic developments into industrial tools, that can substantially advance the analysis and design capabilities of high-end engineering stakeholders. The novel developments and results were achieved using the academic-driven open-source framework \nekpp.
\end{abstract}



%
\section{Introduction}
%
The accurate modelling and prediction of fluid dynamics structures and aerodynamic forces around complex bodies is a classical and challenging problem that lies at the heart of modern automotive design. The increasing availability and lowering cost of computational modelling compared to experimental testing means that computational fluid dynamics (CFD) now forms a key part of the engineering design process. When considering the high Reynolds number flows around the complex bodies found in these problems, there are two key considerations that need to be taken into account: (a) creating highly accurate meshes for the geometry at hand, and (b) producing physically-faithful results, while using numerical discretisations that are necessarily unable to capture all the fluid dynamics scales involved due to the large computational costs required.

In industry, the most common practice revolved around the use of ensemble-averaged steady-state modelling, commonly using either Reynolds-averaged Navier-Stokes (RANS) or, more recently, Detached eddy Simulation (DES). Typically these are used in conjunction with low-order numerical discretisations, predominantly the second-order accurate finite volume method, together with linear meshes of planar-faced elements. Although this approach is computationally cheap, it can undermine the accuracy of the simulations and keep the practitioner from identifying aerodynamic design solutions that would be otherwise feasible if the CFD methodology was able to more faithfully describe the flow physics~\cite{slotnick2014cfd}. Indeed, on one hand, the use of highly diffusive numerical schemes and steady-state modeling strategies leads to suppress under-resolved scales and their feedback on the mean flow, thereby additionally corrupting the accuracy of the simulation. On the other, the use of linear meshes requires high degrees of refinement in order to capture highly-curved surfaces in the geometry, which may introduce artificial diffusion and significantly alter the flow evolution that one would otherwise observe. Addressing point (a) and (b) within the context of industrial applications is therefore of primary importance to advance the state-of-the-art of CFD, and can dramatically enhance the computational aerodynamic analysis and design capabilities of high-end engineering stakeholders. 

Point (a) requires substantial theoretical and algorithmic advances in high-order mesh generation. From this perspective, the aim is to produce a methodology for generating high-order meshes that achieve levels of robustness over a range of very complex cases comparable to that found in commercial linear mesh generators. Indeed, commercial packages, such as Pointwise~\cite{pointwise2017} and Centaur~\cite{centaur2017}, have attempted to add high-order meshing capabilities to their products with variable degree of success. 

The standard approach in generating a high-order mesh is to adopt an \textit{a posteriori} procedure, whereby a coarse linear mesh is first generated, and is then deformed by introducing the geometric curvature into elements lying on the surface. Several research groups have been working on such an approach, see for example \cite{dey2001towards,sherwin2002mesh,luo2004automatic,shephard2005adaptive,sahni2010curved,xie2013generation}, which gives an indication towards the level of interest in this area. 

The challenge of this approach, however, lies in the second step of the process: introducing curvature purely at the surface will typically lead to elements becoming inverted and unsuitable for computation. This has proven a significant stumbling block in the use of high-order methods in industry, and many of the aforementioned works focus on developing techniques to regularize or heal the mesh as part of the \textit{a posteriori} process. While this is the main route to high-order mesh generation, some alternatives exist. These include the use of the Nash theorem to map a Riemannian space onto an isometric Euclidean space of higher dimension~\cite{nash1954c1,zhong2013particle,dassi2014curvature}, semi-structured approaches to block mesh generation~\cite{bucklow2017automated}, anisotropic high-order boundary-layer mesh generation~\cite{moxey2015isoparametric}, and PDE-based approaches based on elastic analogy~\cite{poya2016unified}. 

In this work, we tackle the problem of high-order mesh generation by using a variational framework that uses the minimization of an energy functional. The process begins by defining a linear surface mesh that conforms with the underlying geometry, typically defined in a computer-aided design (CAD) format that follows standard ISO 10303 (informally known as standard for the exchange of product model data, or STEP)~\cite{pratt2001introduction}. The linear surface mesh is then complemented with a linear volume mesh along with a boundary-layer mesh (if required). Finally, the linear surface mesh is made high-order to fit the curved surfaces of the underlying geometry, and the interior mesh entities (edges and surface) of the linear volume mesh are deformed to accommodate the curved boundaries imposed by the geometry. The latter step is achieved by the minimization of the energy functional mentioned above, that can be arbitrarily selected by the practitioner depending on the mesh requirements. The linear mesh is obtained beforehand, and uses a combination of the commercial packages \cf and \ccm for CAD healing and for linear and boundary-layer mesh generation, respectively, as well as \nm~\cite{turner2017thesis}, which is the open-source high-order meshing framework developed as part of the \nekpp project~\cite{nektar++2017,cantwell2015nektar,moxey2020nektar++}.\\

\begin{figure}[htbp]
  \centering
  \label{fig:requirements_cfd}
  \includegraphics[width=0.6\textwidth]{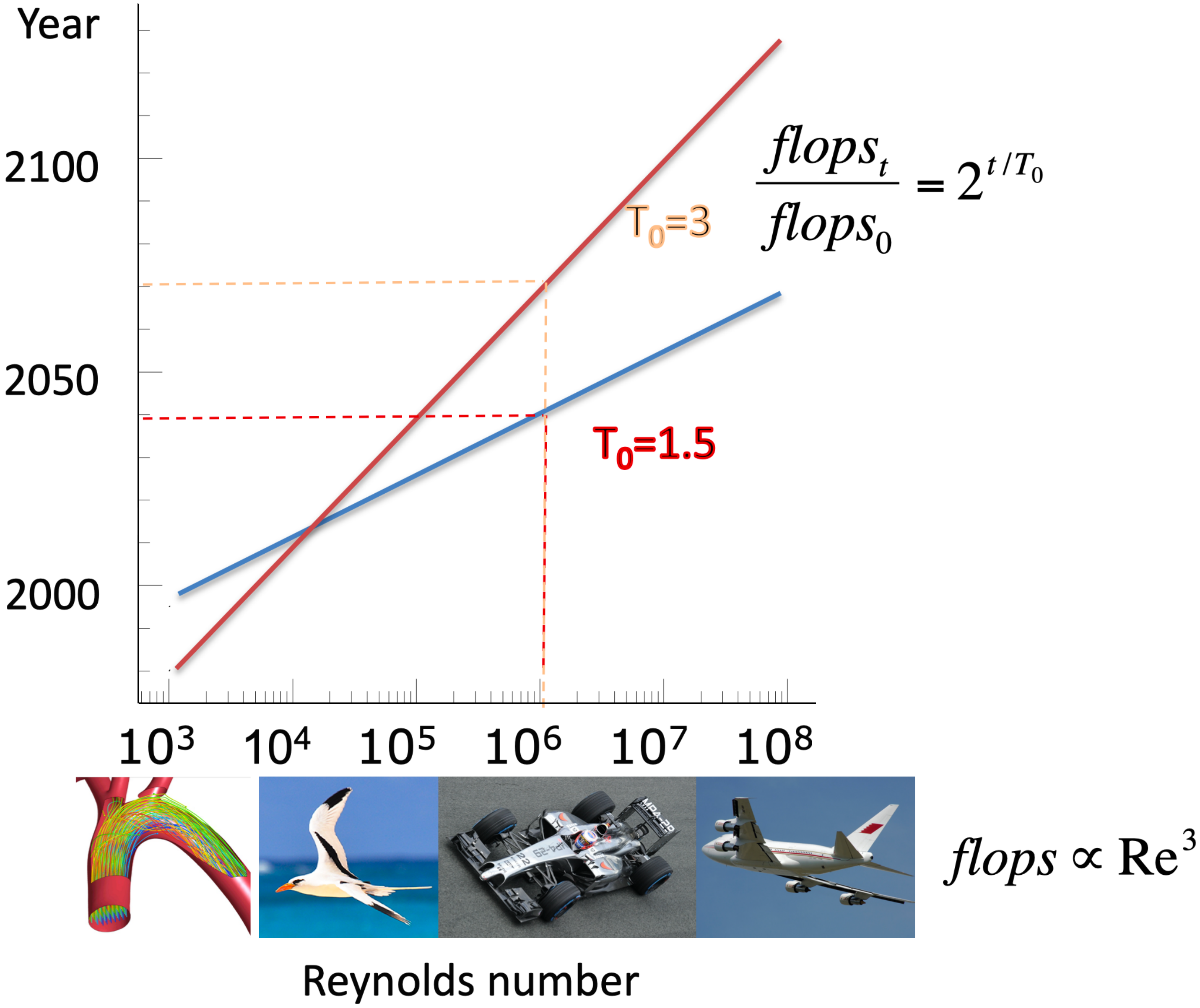}
  \caption{Computational resources required to perform the direct numerical simulation (DNS) as a function of the Reynolds number~\cite{karniadakis1993nodes}. Yellow and red lines represent projections on when DNS capabilities are due to become available as computational power increases following two different Moore law's paths.}
\end{figure}

The second point, (b), is instead closely linked to the impossibility of carrying out simulations that can resolve all the scales present in the problem, that, in turn, leaves part of the flow physics marginally or under-resolved. The computational power required to perform a fully resolved flow simulation, also known as direct numerical simulation (DNS), is a function of the Reynolds number~\cite{karniadakis1993nodes}. From this perspective, we can consider the number of floating point operations per second (FLOPS) that a computer can perform as an approximate measure of computational power, and we can then model the time to solution under different regimes of computational power growths. This gives a rough estimate of the number of years required to achieve DNS capabilities for different Reynolds numbers. \cref{fig:requirements_cfd} depicts two different computational power growths. The red line assumes doubling flops every three years, while the blue line assumes a doubling of FLOPS every one and a half years. They represent a less and more optimistic scenario than Moore's law, respectively. We can appreciate how, under the most optimistic power growth scenario (blue line), DNS capabilities for automotive simulations, which corresponds to Reynolds numbers of the order of $10^6$, will be reached approximately by $2040$. \Cref{fig:requirements_cfd} is an ideal scenario, as it does not take into account additional costs incurred, and it also assumes using the top supercomputers, as soon as they become available. The latter aspect is obviously not achievable for analysis and design purposes in an engineering workflow. Therefore, the ability to perform DNS for industrial purposes at high Reynolds numbers on a regular basis is, at least, two or more decades ahead, even with current advances in computational technologies (e.g.\ many-core architectures, graphical processing units, and quantum computing). Given these limitations, we need to deal with under-resolution for the next several years, while pushing the boundaries of high-Reynolds number flow physics for industrial problems.

A significant body of work has been produced in the past several years to deal with high-Reynolds number problems. As briefly mentioned above, RANS and DES are frequently used in the context of industrial CFD, see for example~\cite{fu2018turbulence,zhang2019turbulence,bush2019recommendations}, but they typically fail to capture unsteady flow features, which may be crucial for the analysis and design of a given aerodynamic solution. This aspect is especially marked when these two modeling choices are adopted in conjunction with low-order numerical schemes, that are commonly characterized by significant diffusion and dispersion, thereby causing an unphysical behaviour of the flow physics \cite{castiglioni2015numerical}. 

In order to overcome these issues, several research groups have been working on LES, both physical (or explicit) and implicit, over the past few decades. Explicit LES uses physical principles such as eddy viscosity to devise closure models that can approximate the under-resolved scales, while implicit LES utilizes the diffusion introduced by the numerical discretization. For a detailed review on LES methods, the interested reader can refer to Pope~\cite{pope2004ten}. LES is particularly promising for automotive applications, since these regimes heavily feature flow physics that steady-state approaches struggle to capture, such as vortex separation and interaction~\cite{slotnick2014cfd}. In addition, when used in conjunction with low-diffusive and dispersive numerical schemes, the resolution power is enhanced and can produce flow simulations at unprecedented accuracy. 

In this work, we use an implicit LES approach, where we treat the under-resolved scales via a novel artificial numerical regularization approach based on a so-called spectral vanishing viscosity (SVV) kernel, and we adopt robust dealiasing techniques to mitigate aliasing driven numerical instabilities that are normally present due to the low-diffusive behavior of the numerical framework. In particular, we show that the use of SVV and dealiasing strategies, along with key improvements in the computational efficiency of kernel calculations, leads to a robust (in terms of numerical stability) and accurate high-fidelity LES framework for the analysis and design of complex industrial problems. This is showcased in the successful high-fidelity simulation of the real geometry of the Rp1 automotive car at a Reynolds number of $10^6$ based on the car length. All the simulations performed use the incompressible Navier-Stokes solver implemented in the open-source spectral/$hp$ element framework \nekpp. The solver implements an operator splitting scheme \cite{karniadakis1991high,guermond2003velocity}, where the spatial discretization is achieved via a continuous Galerkin method \cite{karniadakis2005spectral}, and the time-discretization is obtained via a second-order implicit-explicit time-stepping strategy \cite{vos2011generic}. The numerical simulations performed were achieved with hybrid meshes containing both prisms (to discretize the boundary-layer regions in proximity to the body of the car) and tethraedra (to discretize the farfield of the domain).

In this paper, we present the key advancement in high-order meshing and simulation technologies that we implemented in \nm and \nekpp in order to achieve the challenging simulation of the Rp1 car, with the aim translating academic research efforts to industrial CFD applications. In particular, section~\ref{sec:mesh} describes the high-order meshing strategy we adopted, focusing on the steps required for surface and linear mesh generation as well as on the \textit{a posteriori} high-order deformation. In section~\ref{sec:num-met}, we describe the novel simulation technologies implemented in terms of regularisation approaches for the under-resolved scales, dealiasing techniques for numerical stability, and computational efficiency. Section~\ref{sec:car} presents the high-fidelity simulation of the Rp1 car, and finally, in section~\ref{sec:conclusions}, we draw some final remarks.

%
\section{High-order mesh generation for complex geometries}\label{sec:mesh} 
%

The starting point and essential prerequisite for any high-fidelity LES, within either a finite element or finite volume framework, is the generation of a geometrically-accurate and high-quality mesh. Mesh generation is a complex and involved process, particularly when the geometry under consideration is a complex, three-dimensional body, since the important regions that need to be resolved within the domain are often unknown \emph{a priori}. The process may therefore require many iterations of meshing, simulation and analysis in order to arrive at a final mesh deemed suitable enough for accurate results. 

Mesh generation tools are generally required to be both flexible and robust, leading to a wide variety of approaches. In industry, the majority of mesh generation strategies lead to a linear mesh, as this is what most simulation packages are capable of supporting. While widely adopted, the use of linear meshes could alter the flow physics in proximity of curved geometrical features, that in turn can lead to a critical reduction in simulation accuracy. The use of higher-order methods can address this loss of accuracy, as one generates boundary-conforming, curvilinear meshes which describe more accurately the underlying geometry for the purposes of simulation.

However, the generation of such meshes in a robust and efficient manner for a realistic three-dimensional geometry remains an open and challenging problem. In fact, the introduction of additional curvature within elements that are connected to the geometry -- and that must conform to the boundary -- will often lead to significant reduction in the quality of the mesh, or indeed deform the element to such an extent that it is self-intersecting and thus unsuitable for simulation usage. For simulations in fluid dynamics, where boundary layers of highly anisotropic elements are required to resolve the shear of the flow near the boundary, this issue becomes even more pronounced. Finding routes to mitigate and address these issues is the central challenge for the high-order mesh generation community, and forms a key bottleneck to tackle in the wider adoption of high-order methods. In this section, we outline our process to try to overcome this issue, and highlight the process within the context of realistic automotive geometries.

\subsection{Approaches to high-order \emph{a posteriori} mesh generation}

Our framework to produce a viable high-order mesh, similar to other state-of-the-art methods \cite{dey2001towards,sherwin2002mesh,sahni2010curved,hindenlang2010unstructured,xie2013generation}, uses an \textit{a posteriori} approach. In this setting, one first generates a coarse, linear mesh of the desired geometry. From this, high-order nodes are added within each element. Any nodes that lie on curved surfaces are projected onto the surface in such a manner as to enforce a boundary-conformal representation. In order to achieve the curvature required by the geometry being meshed, we typically adopt an isoparametric approach, in which the same basis functions that are used to represent high-order solutions within each element are also used to define the geometric coordinates within the element. Such a mapping can be expressed therefore as the summation
\begin{equation}
\boldsymbol{x} = \boldsymbol{\phi}_{M}(\boldsymbol{\xi}) = \sum_{n=1}^{N}\boldsymbol{x}^{n}\boldsymbol{\ell}_{n}(\boldsymbol{\xi}) \, ,
\label{eq:interp}
\end{equation}
where $\boldsymbol{x}$ is the spatial location of the a point within an element, $\boldsymbol{\xi}$ is a corresonding location within a reference element, $\boldsymbol{x}^{n}$ is the location of a given node in the element, and $\boldsymbol{\ell}_{n}$ is the polynomial interpolant (e.g. a Lagrange polynomial interpolant). In particular, mesh entities, namely faces and edges, on the boundary of the geometry are curved by adding high-order nodes, and then enforcing that the nodes lie on the boundary. Moreover, the position of the nodes $\boldsymbol{x}^n$ must be adjusted so that the resulting mapping is a faithful polynomial representation of the geometry~\cite{sherwin2002mesh}.

\begin{figure}[htbp]
  \centering
  \label{fig:curvings}
  \includegraphics[width=0.99\textwidth]{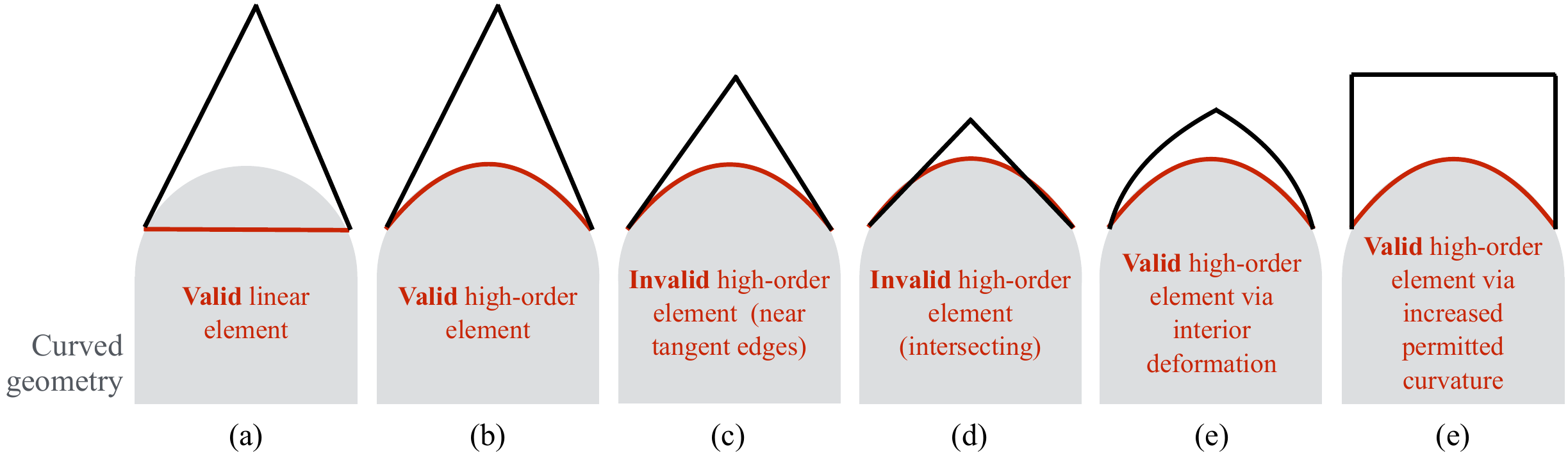}
  \caption{Valid and invalid high-order elements.}
\end{figure}

Although from this description this may seem a simple task, for complex geometries that can have several hundred curved surfaces, the ability to create valid high-order elements (that is elements of order greater than 2) is extremely challenging. In many cases, when simply curving the edge and/or face in proximity to a given curved geometrical feature, it is relatively easy to generate invalid elements. Figure~\ref{fig:curvings} shows seven mesh elements in proximity of a curved surface, depicted in grey. The first mesh element (a) is a valid linear element that completely misses the curvature of the underlying surface. Indeed, in order to capture the curvature of the surface with a linear mesh it is necessary to add more elements around the surface than would be necessary with high-order elements; the latter benefits from a superior accuracy that allows for larger element sizes than linear elements. The second element (b) is a valid high-order element, where only the edge in proximity to the geometry is curved. The third and fourth elements, (c) and (d), are both invalid high-order elements, due to near tangent (c) or intersecting edges (d). Mesh elements (b), (c), and (d), clearly highlight the challenge of high-order mesh generation. The curvature of the geometry dictates the size of the element one can have, and in several occasions, it could lead to invalid elements. Mesh elements (e) and (f) show two possible solutions to the invalid elements (c) and (d). Mesh element (e) deforms the interior edges to accommodate the curvature of the boundary, while (f) uses a quadrilateral element to increase the permitted curvature. In both cases, there are no longer near tangent or intersecting edges. 
From the examples in figure~\ref{fig:curvings}, it should be immediately clear that the challenge of high-order mesh generation is multifold. On the one hand, it is necessary to create curved mesh entities that conform to the underlying curved geometrical features (where the geometry is typically provided in a CAD format). On the other hand, it is essential to develop a strategy to handle invalid high-order elements and produce a valid high-quality mesh. 

Our high-order meshing strategy handles the challenges just described and consists of three steps, (1) the definition of the geometry in such a way that is can be efficiently queried for curvilinear mesh generation, (2) the generation of a high-quality linear mesh, and (3) the generation of curved elements. These three crucial steps are described in detail in sections~\ref{subsec:geometry-def}, \ref{subsec:linear-mesh} and \ref{subsec:high-order-elements}, respectively. The corresponding meshing framework has been implemented in the code \nm, that is part of the open-source \nekpp library~\cite{cantwell2015nektar}. For a more comprehensive explanation of the mesh generation process, the interested reader can refer to Turner et al.~\cite{turner2018}.

\subsection{Geometry definition}\label{subsec:geometry-def}

The first stage of our meshing approach is the definition of the geometry in a way that ensures its accurate representation and permits geometrical enquiries during the various mesh-generation stages. The most common approach to define an arbitrary complex geometry is by a boundary representation (also referred to as ``BRep''). The boundary representation contains surface patches and curves that link these surface patches. The surface patches and curves defining the underlying geometry are typically topologically connected, thereby forming a closed shell. These surfaces and curves are parametrised onto a lower-dimensional space, that is the one-dimensional space $\boldsymbol{x}(s)$ for curves, and the two-dimensional space $\boldsymbol{x}(s_{1},s_{2})$ for surfaces, where $\boldsymbol{x} = (x,y,z)$ denotes the three-dimensional coordinate of a point in the boundary representation of the geometry. The main geometrical enquiries required for mesh generation consists of three key operations: 
\begin{itemize}
\item parametric coordinate to real space location, i.e. determining $\boldsymbol{x}$ from $(s_1,s_2)$;
\item projection of a location in space onto a boundary representation entity, i.e. determining $(s_1,s_2)$ from $\boldsymbol{x}$; and 
\item calculation of the first- and second-order derivatives of the parametric
mapping with respect to $s_1$ and $s_2$.
\end{itemize}
In addition, in this work we do not consider the effects of incomplete or low-quality geometric representations; The boundary representation must be watertight and should represent accurately the geometry under investigation. These two requirements are crucial for high-order mesh generation, as they impact the number of high-order valid elements that is possible to generate and the quality of the mesh.

The assumption adopted in this work is that the geometry is given in a standard CAD (computer-aided design) format that supports a boundary representation description, such as STEP. In several CFD applications, this assumption is fairly common, especially among industry practitioners. The CAD object is then probed via \nm, that performs geometrical queries through interfaces to a CAD engine. To give support for different CAD engines and implementations, \nm implements a lightweight interface to the key operations above. Concrete implementations of this use CADfix~\cite{CADfix} and OpenCASCADE~\cite{opencascade}, both of which permit reading CAD data in various formats, including STEP. In particular, the checks and queries on the CAD object were performed by \nm, while the correction of any errors was performed by CADfix which provides an extensive toolset for fixing geometrical inconsistencies in CAD data. CADfix is a commercial software, that provides extensive CAD handling tools, useful for healing and modifying the underlying CAD geometry definition. It also provides a linear mesh generator based on the medial approach that is useful for decomposing the domain into partitions. This decomposition has proven useful for designing high-quality boundary-layer meshes near the wall surfaces, aspect that was critical for the high-fidelity simulations in section~\ref{sec:car}. OpenCASCADE is an open-source code with reduced capabilities than CADfix, and represents the default CAD handler within \nm, due to its free-availability, that is at the core foundations of the \nm and \nekpp projects. Nevertheless, the use of CADfix in this work was essential to obtain a robust and efficient high-order meshing pipeline, and ultimately obtaining valid high-quality meshes.  Access to CADfix, or some alternative approach to geometry healing and defeaturing, is a fundamental requirement for the development of a robust and successful meshing strategy.

Once the CAD representation of the geometry is satisfactory, it is possible to complete the definition of the computational domain, including the external boundaries, and proceed to the next meshing stage, that is generating a valid linear mesh. This step is described next.

%
\subsection{Linear mesh}\label{subsec:linear-mesh}
%

The generation of a high-quality linear mesh from the CAD representation of the geometry is a demanding task. Indeed, it is necessary to have a sufficiently coarse initial mesh that serves as a starting point for the high-order meshing process. This initial mesh is typically obtained via a user-driven manipulation of the mesh spacing definition at all points within the computational domain to achieve satisfying geometric accuracy, coarseness and mesh gradation. This procedure is typically time-consuming and challenging, possibly involving a large number of iterations before obtaining a linear mesh that satisfies the quality required by the practitioner.

In order to automate this step as much as possible, we combined the advanced linear meshing capabilities of the commercial package \ccm with the high-order meshing procedures of \nm. This allowed us to produce an automatic industrial linear meshing pipeline, that required minimal intervention from the user. The main obstacle that we encountered when using \ccm, and that is common to encounter when using commercial CFD software, was the lack of information relating the linear mesh to the original CAD definition. Often the CAD surfaces are replaced by a fine triangulation that represents the ``true'' boundary of the geometry within a given tolerance. The linear mesh is then built using this approximate representation and the original CAD information discarded. This is a significant issue since inaccuracies in the definition of the geometry could lead to errors in the high-order mesh generation process, that in turn could lead to inaccurate flow simulation results. 

In order to tackle this problem, we used the projection of the linear mesh onto the CAD surfaces to reconstruct the missing CAD information. In particular, we imported the linear mesh generated by \ccm into \nm. The import was performed by incorporating curvature-driven mesh size restrictions such that the boundary of the surface mesh is a reasonably close representation of the underlying CAD model. Once the import of the linear mesh into \nm is complete, the CAD model is processed in order to obtain parametric coordinates in the CAD representation that will be used to enhance the adherence of the linear mesh to the CAD model. From this perspective, the first step is to linearise each surface of the CAD model via an ``auxiliary'' triangulation, so that each CAD surface has two representations, the CAD model itself, and the ``auxiliary triangulation''. The second step is to define a bounding box (inflated by 5\% in each direction), for each CAD surface, and store these boxes into a $k$-dimensional ($k-d$) tree data structure~\cite{bentley1975}. Once the two CAD representations are available and the bounding boxes are defined for each CAD surface, the surface mesh nodes are processed to obtain their CAD parametric coordinates. More specifically, we associate each node of the surface mesh that was generated in \ccm to a list of potential CAD parent surfaces by querying the bounding box $k-d$ tree. If a node is within a bounding box, the surface is added to the list of potential parents surfaces. The use the $k-d$ tree makes this process particularly efficient and reduces the number of potential parent surfaces to just a few candidates, typically on the order of 2 or 3.  The surface mesh node is then projected onto the CAD surface for each of the parent surface candidates. This is achieved by identifying the nearest vertex in the ``auxiliary triangulation'' of the CAD surface, and then using this vertex as an initial guess for the nonlinear optimization problem associated to node-to-surface projection. The parent surface is finally identified as the closest surface to the projected node.

We note that, since the ``auxiliary triangulation'' of the CAD surfaces is not directly related to the \ccm linear mesh, the surface node of the linear mesh might not lie on the CAD surface. Therefore, once the parent CAD surface associated to the mesh node is identified, the mesh node is moved to the CAD surface it belongs to, thereby improving the adherence of the linear mesh to the underlying CAD surfaces. The node movement just described is always performed, except in two cases: if the required displacement of the node is greater than 10\% of the length of the edges in the local mesh, and if the movement of the node induces inverted (i.e.\ invalid) elements. In these two instances, the node is added to an ``exclusion set'' with a view to preventing deterioration of mesh quality or generation of invalid elements.
The final product of the projection process just described is a set of nodes that are accurately placed onto the original CAD surfaces, and a set of nodes that are placed into an ``exclusion set''. Each node, regardless if in the ``exclusion set'' or not, will have all the parametric information required to generate the final high-order curving of the surface mesh, with the exception that any mesh entity (edge or face) that has a node in the ``exclusion set'' will be left linear. The approach outlined has been named \textit{projection curving}~\cite{turner2017thesis}, as it involves the projection of the surface mesh generated via \ccm onto an ``auxiliary triangulation'' of the CAD surfaces. This approach is the one used in the final industrial meshing pipeline, that was adopted for the last design stage, D3, of the road car in section~\ref{sec:car}. 
A boundary-layer mesh was also created by \ccm, and consisted of a prism layer in proximity of the surfaces of the geometry. 

We finally point out that we initially attempted two additional alternatives, which were however discarded because they became too time-consuming and user-driven. These are named \textit{analytic curving}, and were used for the first two designs, D1 and D2, of the road car presented in section~\ref{sec:car}. In both the \textit{analytic curving} strategies, all the parametric information associated with the linear surface mesh was known, hence high-order curving of the surface was a relatively simple task, unlike in the \textit{projection curving} approach. The first \textit{analytic curving} strategy used the linear meshing capabilities of \nm to create the surface mesh that was later exported to \ccm for the generation of the volume mesh, and of the boundary layer mesh (constituted by a near-wall macro prism layer). This approach was however extremely time-consuming as it involved dozens of iterations between \nm, \ccm, and the original CAD model. The second \textit{analytic curving} strategy tried to address the bottlenecks of the first by allowing the \ccm linear mesh generator to create the surface mesh from a linearised CAD surface obtained using \nm. The surface mesh was then forced to explicitly conform to the boundaries of the CAD surfaces. This is critical in an industrial meshing pipeline, as it is typically impractical to clean complex CAD models with thousands of surfaces, so as to guarantee a clean enough CAD to enforce the surface mesh to obey to the CAD surfaces.

\subsection{Generation of high-order elements}\label{subsec:high-order-elements}

The third stage is the generation of the curved surfaces, where the coarse linear mesh generated in the previous step and described in section~\ref{subsec:linear-mesh} is deformed to accommodate the curvature of the boundary. We already mentioned in section~\ref{subsec:linear-mesh}, that the creation of curved boundary mesh entities that conform to curved geometric features can be handled via two different strategies, \textit{projection curving} and \textit{analytic curving}, and we found the former to be the optimal approach for complex industrial CAD models. While curving the boundaries attached to the geometry is a relatively simple task once the parametric coordinates of the surface mesh are known and accurate, the creation of valid high-order elements remains a challenge, as depicted in figure~\ref{fig:curvings}. Indeed, one of the key research themes in \textit{a posteriori} high-order mesh generation is to find suitable deformations of the interior mesh entities. The ability to deform interior mesh entities allows for better mesh quality and for a significant reduction of invalid elements, see figure~\ref{fig:curvings}-(e). To obtain suitable deformation of interior mesh entities, there exist a certain number of alternatives, including methods based on partial differential equations~\cite{persson2009curved,xie2013generation,moxey2014thermo,fortunato2016high,hartmann2016generation} and on the minimization of an energy functional~\cite{toulorge2013robust,gargallo2015optimization}. In this work we pursued the latter, and developed a variational framework able to encapsulate arbitrary energy functionals in a unified high-order meshing framework that has been implemented in \nm. 

The variational formulation begins with a coarse linear mesh that was created as described in section~\ref{subsec:linear-mesh}. Following \cite{turner2017thesis}, we consider a linear mesh with $N_{el}$ straight-sided elements, that is $\Omega_{I} = \bigcup_{e=1}^{N_{el}} \Omega_{I}^{e}$, where each element is provided with a high-order polynomial basis function based on standard Lagrange polynomials. The mapping between the straight-sided mesh, $\Omega_{I}$, and the curvilinear mesh, $\Omega$, is denoted by $\phi : \Omega_{I} \rightarrow \Omega$.

\begin{figure}
  \centering
  \label{fig:deformation}
  \includegraphics[width=0.60\textwidth]{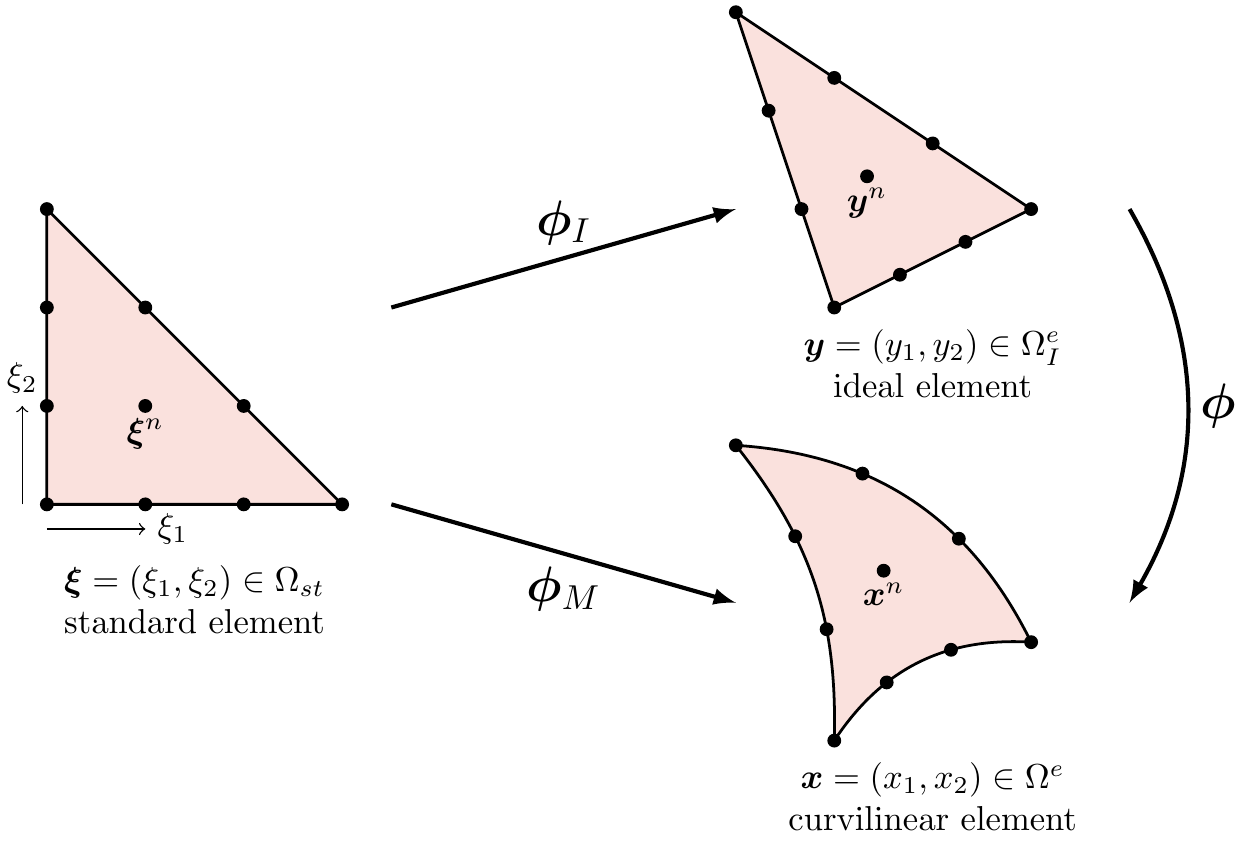}
  \caption{Deformation of a straight-sided triangular element, referred to as standard element, onto an ideal element via mapping $\phi_{I}: \Omega_{st} \rightarrow \Omega^{e}_{I}$, and onto a curvilinear element via mapping $\phi_{M}: \Omega_{st} \rightarrow \Omega^{e}$. The mapping of the deformation $\phi$ from the ideal element and the curvilinear element is defined as the composition of the two mappings, $\phi:\phi_{M} \circ \phi_{I}^{-1}$. Figure taken from Turner et al.~\cite{turner2018}}
\end{figure}

With reference to figure~\ref{fig:deformation}, we denote the coordinates of the standard (also referred to as reference) element as $\boldsymbol{\xi} \in \Omega_{st}$, those of the curvilinear element as $\boldsymbol{x} \in \Omega^{e}$, and those of the ideal element as $\boldsymbol{y} \in \Omega^{e}_{I}$. The mappings between the elements in figure~\ref{fig:deformation} are constructed in an isoparametric fashion, such that the nodes $\boldsymbol{\xi}^{n}$ defining the Lagrange basis functions on the standard element map to $\boldsymbol{y}^{n}$ under $\phi_{I}$ and to $\boldsymbol{x}^{n}$ under $\phi_{M}$. The example reported in figure~\ref{fig:deformation} is for triangular elements but it can be readily generalized to any element shape, in both two and three dimensions.

Once the nomenclature for the mesh is specified, we can define the energy functional adopted throughout this work, that is 
\begin{equation}
\mathcal{E}(\nabla\phi) = \int_{\Omega_{I}} W(\nabla\phi)\text{d}\boldsymbol{y} \,,
\label{eq:functional}
\end{equation}
where $W$ depends on the deformation gradient tensor $\nabla\boldsymbol{\phi}(\boldsymbol{y}) = \frac{\partial \boldsymbol{\phi}}{\partial \boldsymbol{y}}$ and its determinant $J = \text{det}[\nabla\boldsymbol{\phi}]$, also referred to as the Jacobian.
The functional $W$ can assume various forms depending on the associated energy adopted. We have implemented functionals associated to linear elasticity energy, isotropic hyperelasticity energy, Winslow equation energy, and a distortion-based energy. Details of these functionals can be found in Turner et al.~\cite{turner2018} and in Turner~\cite{turner2017thesis}, and are not reported here for the sake of brevity.

Since the ideal mesh is a union of many elements, the functional can subsequently be defined element by element
\begin{equation}
\mathcal{E}(\nabla\phi) = \sum_{e=1}^{N_{el}}\int_{\Omega^{e}_{I}} W(\nabla\phi)\text{d}\boldsymbol{y} = \sum_{e=1}^{N_{el}} W[\nabla\phi_{M}(\boldsymbol{\xi})\nabla\boldsymbol{\phi}_{I}^{-1}(\boldsymbol{\phi}_{I}(\boldsymbol{\xi}))]\, \text{det}(\nabla\boldsymbol{\phi}_{I}) \text{d}\boldsymbol{\xi}\,,
\label{eq:functional-element}
\end{equation}
where we used the composition $\boldsymbol{\phi} = \boldsymbol{\phi}_{M} \circ \boldsymbol{\phi}_{I}^{-1}$. To fully define equation~\eqref{eq:functional-element}, we need to specify the expression of $\boldsymbol{\phi}_{M}$ and $\boldsymbol{\phi}_{I}$. The first, $\boldsymbol{\phi}_{M}$, assumes the usual form of an isoparametric mapping commonly found in nodal spectral element methods and specified in equation~\eqref{eq:interp}. The second, $\boldsymbol{\phi}_{I}$, assumes an analytical form that is a combination of linear finite element basis functions and that depends on the element shape. For instance, in the case of a triangle with vertices $\boldsymbol{v}^{1}$, $\boldsymbol{v}^{2}$, and $\boldsymbol{v}^{3}$, the mapping is given by
\begin{equation}
\boldsymbol{\phi}_{I} = (\boldsymbol{v}^{2} - \boldsymbol{v}^{1})\xi_{1} + (\boldsymbol{v}^{3} - \boldsymbol{v}^{1})\xi_{2} + \boldsymbol{v}^{1}\,.
\label{eq:triangle}
\end{equation}

The minimization of the functional is then formally defined as follows 
\begin{equation}
\text{find}\,\min_{\boldsymbol{\phi}} \mathcal{E}(\boldsymbol{\phi}) = \min_{\boldsymbol{\phi}} \int_{\Omega_{I}}W(\nabla\boldsymbol{\phi})\text{d}\boldsymbol{y}\,,
\label{eq:minimization}
\end{equation}
where we seek the configuration of nodes of the linear mesh, denoted by $\boldsymbol{N} = [\boldsymbol{n}^{1}, \dots, \boldsymbol{n}^{N}_{\text{nodes}}]$, that minimizes the energy functional in equation~\eqref{eq:minimization}. In practice, this means to start from the initial distribution of nodes in the linear mesh, $\boldsymbol{N}^{0}$, in which the curvature of the boundary mesh entities is imposed, regardless of whether it causes self-intersections. Then, the minimization proceeds in iteratively updating $\boldsymbol{N}$ towards lower energy configurations, that is: $\boldsymbol{N}^{k} \rightarrow \boldsymbol{N}^{k+1}$, such that $\mathcal{E}^{k} \rightarrow \mathcal{E}^{k+1} < \mathcal{E}^{k}$. The minimization is achieved via a gradient- and Hessian-based method, and utilizes a relaxation strategy, whereby a local optimization problem is solved in place of a global one. More specifically, the minimization in equation~\eqref{eq:minimization} is performed for a subset of elements that are influenced by a change in the position of the node under consideration. This requires the solution of a smaller and less computationally expensive problems, compared to the global problem that would be otherwise necessary to solve. The minimization of equation~\eqref{eq:minimization} can then be recast into the minimization of a set of energies, each of which belongs to a node and the subset of elements the node influences. The stopping criterion used to find the global minimum based on the set of local energies, is $||\boldsymbol{N}^{k+1} - \boldsymbol{N}^{k}||_{\infty} / L < \epsilon$, where we used $\epsilon = 10^{-6}$, and $|| s ||_{\infty} = \max\{|s_{1}|, \dots, |s_{n}|\}$ is the uniform norm.

This process also requires the computation of the gradient of $\mathcal{E}(\boldsymbol{\phi})$, with respect to the movement of node positions. As a first approach, these gradients can be approximated by use of finite difference stencils. However, our testing indicates that this approach is both expensive and sensitive to the size of the stencil. Instead, we make use of an analytic expression for the desired gradients. Ultimately, applications of the chain rule therefore require the same derivatives for $\boldsymbol{\phi}_I$ and $\boldsymbol{\phi}_M$. Although the ideal mapping $\boldsymbol{\phi}_{I}$ readily admits an analytic expression, more care is required for $\boldsymbol{\phi}_{M}$, but by exploiting properties of the Lagrangian expansion~\eqref{eq:interp}, analytic forms of the derivatives can be obtained, which are given in the appendix of~\cite{turner2018}.

\begin{figure}[h!]
\centering 
\includegraphics[keepaspectratio=true, trim={0 0 0 0}, clip, width=0.65\textwidth]{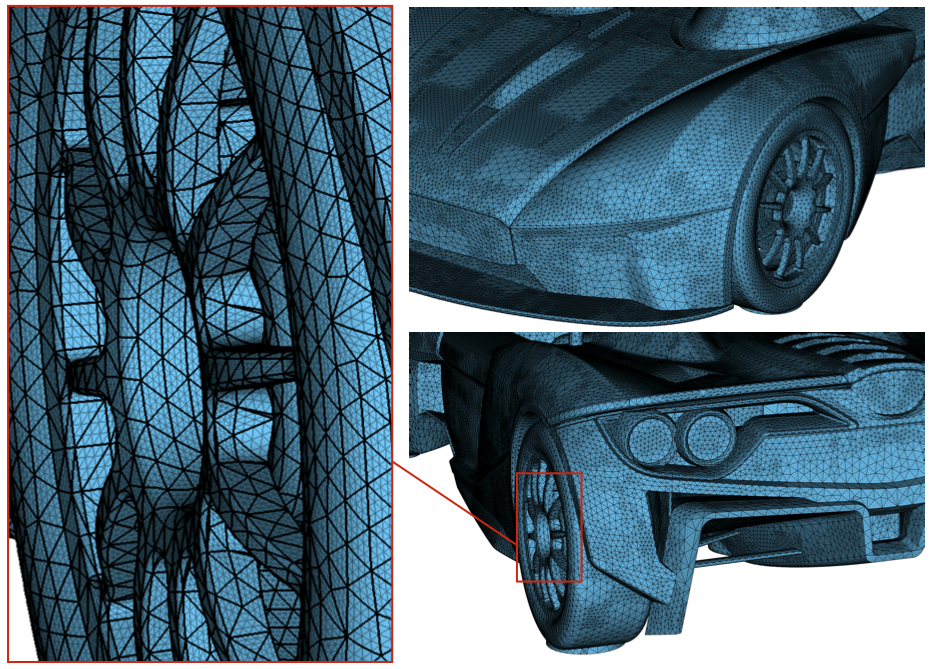}
\caption{Details of the Rp1 high-order mesh produced using the high-order meshing workflow presented in section~\ref{sec:mesh}. We note that the spokes in the wheel were not included in the final simulations to reduce complexity.}
\label{fig:rp1-spokes}
\end{figure}

Figure~\ref{fig:rp1-spokes} shows the high-order surface mesh of the Rp1 car, resulting from our meshing process. It is possible to appreciate the high-order elements tessellating the CAD surface. We note that the spokes within the wheel shown in figure~\ref{fig:rp1-spokes} were not included in the simulations presented in section~\ref{sec:car} to avoid the requirement for a moving mesh.

As a final remark, we point out that, all those nodes on the boundary surface mesh that belonged to the ``exclusion set'' are left straight-sided.  This leads to a small but not insignificant proportion of the surface mesh entities that will be straight-sided as opposed to curved, and could have significant impact on the geometric accuracy of the mesh and hence the solution. However, in the results that we present in section~\ref{sec:car},  we observe little noticeable impact, since often these excluded elements were in locations where multiple CAD surfaces met but not typically in 
places of interest to the flow physics. Obviously this cannot be always guaranteed to be true. A full investigation into this has not been conducted yet, but this work proves that high-order meshes can be produced on industrial models and sensible results obtained with little meshing effort. Indeed the final projection meshing pipeline only required one execution of the linear mesh generation and \nm, with no need for repetition with either system or the CAD.

%
\section{Numerical methodology} \label{sec:num-met}
%

The model underlying the simulations carried out in this study is constituted by the incompressible Navier-Stokes equations for a Newtonian fluid 
\begin{subequations}
\begin{equation}
\frac{\partial \mathbf{u}}{\partial t} + \mathbf{u} \cdot \nabla \mathbf{u} = -\nabla p + \nu \nabla^2 \mathbf{u} +  \mathbf{f} 
\label{eq:momentum}
\end{equation}
\vspace{-0.75cm}
\begin{equation}
\nabla \cdot \mathbf{u} = 0 \, ,
\label{eq:mass}
\end{equation}
\label{eq:ns}
\end{subequations}
where $\mathbf{u} = (u,v,w)$ is the velocity vector, $p$ denotes the pressure, $\nu$ is the kinematic viscosity, and $\mathbf{f}$ is a generic forcing term. This set of equations are particularly suited to accurately model the flow around a road car, as compressibility effects are deemed to be unimportant at the speeds reached. For diffusers and other heated parts of the car, the compressible Navier-Stokes equations are better suited. This was however out of the scope of the current investigation, which was mainly focussed on the external aerodynamics of the car. 

The system of equations~\eqref{eq:ns} and its numerical discretization has been implemented in the open-source spectral/$hp$ element library \nekpp. The equations are solved via an operator splitting scheme (sometimes referred to as a velocity correction scheme)  \cite{karniadakis1991high,guermond2003velocity} in combination with a consistent boundary condition for the pressure Poisson equation. In particular, following \cite{guermond2003velocity}, we recast problem~\eqref{eq:ns} into a weak pressure Poisson problem by taking the inner product over the solution domain $\Omega$ of equation~\eqref{eq:momentum} with respect to the gradient of the test function, $\nabla v$, i.e.
\begin{equation}
 \int_{\Omega} \nabla v \cdot \frac{\partial \mathbf{u}}{\partial t}  + 
 \int_{\Omega} \nabla v \cdot {\mathcal {N}}(\mathbf{u}) = -\int_{\Omega} \nabla v \cdot \nabla p  +  \int_{\Omega} \nabla v \cdot \nu \nabla^2 \mathbf{u} \, ,
 \label{eqn.weakp}
\end{equation}
where $\mathcal{N}(\mathbf{u}) = \mathbf{u} \cdot \nabla\mathbf{u}$ and we set the forcing term $\mathbf{f}$ to zero.  Using the identity $\nabla^2 \mathbf{u} = - \nabla \times \nabla \times \mathbf{u} + \nabla (\nabla \cdot \mathbf{u})$, we can enforce the divergence to be zero by setting the last term to zero. By integrating the first, second and last term in equation~\eqref{eqn.weakp} by parts we get the weak pressure equation
\begin{align}
  \int_{\Omega} \nabla v \cdot \nabla p^{n+1}  &=  \int_{\Omega} v\, \nabla \cdot \left ( \frac{\partial \mathbf{u}}{\partial t}^{n+1}  + \mathcal{N}(\mathbf{u})^{n+1} \right )  \nonumber \\
  &-  \int_{\partial \Omega} v  \left ( \frac{\partial \mathbf{u}}{\partial t}^{n+1}  + \mathcal{N}(\mathbf{u})^{n+1} + \nu \nabla \times \nabla \times \mathbf{u}^{n+1} \right ) \cdot \mathbf{n}\, ,
 \label{eqn.weakp1}
\end{align}
where $\partial \Omega$ is the boundary of the domain, $\mathbf{n}$ indicates the outward pointing normal to the domain $\Omega$ and the superscript $(n+1)$ denotes the time step that aim to compute. Note that in equation~\eqref{eqn.weakp1} we have also used the identity $\nabla \cdot (\nabla \times \nabla \times \mathbf{u}) = 0$. By using a backward approximation of the time derivative,
\begin{equation}
  \frac{\partial \mathbf{u}}{\partial t}^{n+1} \simeq \frac{\gamma_0 \tilde{\mathbf{u}}^{n+1} - \hat{\mathbf{u}}}{\Delta t}
  \label{eqn:velref}
\end{equation}
along with a consistent extrapolation for the non-linear terms $\mathcal{N}^{*,n+1}$, where $\tilde{\mathbf{u}}^{n+1}$ is an intermediate velocity, $\gamma_0$ is a constant, and $\hat{\mathbf{u}}$ is an extrapolation of the velocity, we can finally calculate the pressure at step $n+1$. We can then solve for the velocity at step $n+1$ by using the pressure at $n+1$ just calculated. This leads to solving the following Helmholtz problem
\begin{equation}
  \Bigl(\Delta-\frac{\gamma_0}{\nu \Delta t}\Bigr)\mathbf{u}^{n+1}=-\Bigl(\frac{\gamma_0}{\nu \Delta t}\Bigr)\hat{\mathbf{u}} + \frac{1}{\nu} \nabla p^{n+1}\, .
  \label{eqn:helmholtz}
\end{equation}
The velocity correction scheme just outlined (equations~(\ref{eqn.weakp1}),(\ref{eqn:helmholtz})) is spatially discretized using a high-order continuous Galerkin formulation \cite{karniadakis2005spectral}. Here, one typically expresses the solution in terms of global basis functions $\Phi_k(\mathbf{x})$ as follows 
\begin{equation}
\mathbf{u} \approx \mathbf{u}_{h} = \sum_{k} \mathbf{u}_{k}(t)\Phi_k(\mathbf{x})\, ,
\label{eq:expansion}
\end{equation}
where $\mathbf{u}_k$ are the weights or coefficients of the polynomial expansion, and we introduced the superscript $\delta$ to denote the spatial discretization of $\mathbf{u}$. Equation~\eqref{eq:expansion} can be decomposed into local boundary and interior function, such that equations\eqref{eqn.weakp1} and \eqref{eqn:helmholtz} can be written in terms of element-wise contributions (omitted here for the sake of brevity). The elemental polynomial expansions can be written as 
\begin{equation}
\mathbf{u}_{\Omega_{e}} \approx \mathbf{u}^{\delta}_{\Omega_{e}} = \sum_{j=0}^{P} \ell_{k}(t)\phi_k(\mathbf{x})\, ,
\label{eq:elemental-expansion}
\end{equation}
where the elemental basis functions $\phi_k(\mathbf{x})$ used are \textit{modified Jacobi} or \textit{Lagrange} polynomials defined on a set of Gauss-Lobatto-Legendre quadrature points, $\ell_{k}$ are the elemental coefficients of the expansion, and we introduced the element $\Omega_e$ of our domain discretization. The global-wise continuity of the solution required by the continuous Galerkin method is then enforced by an assembly procedure. The interested reader can refer to \cite{karniadakis1991high,guermond2003velocity} as well as to \cite{karniadakis2005spectral} (Chapter 4 and 8). The semi-discrete form of the incompressible Navier-Stokes equations that is obtained using the velocity correction scheme and the continuous Galerkin method as spatial discretization and that was just introduced is advanced in time via a second-order implicit-explicit time-stepping strategy \cite{vos2011generic}. 

In order to successfully achieve the challenging high-Reynolds number LES described in section~\ref{sec:car}, we needed to significantly enhance the numerics implemented in \nekpp, especially in terms of (i) numerical stability, (ii) regularization techniques for under-resolved scales, and (iii) computational efficiency. These three key building blocks are discussed in the following.

\subsection{Numerical stability}\label{subsec:dealiasing}

Numerical stability was a key aspect that we needed to carefully take into account in order to successfully simulate the road car presented in section~\ref{sec:car}, at the Reynolds numbers required by industry. To achieve this, we  adopted  polynomial dealiasing via consistent integration~\cite{mengaldo2015dealiasing}, combined with spectral vanishing viscosity (SVV) \cite{karamanos2000aspectral,kirby2006stabilisation,moura2020spatial}. The former helps suppressing aliasing-driven instabilities that are due to inexact numerical integration of the equations, while the latter targets all those flow scales that are necessarily under-resolved in high-Reynolds number simulations. While SVV primary target was not the numerical stability of the simulations, it was necessary the use of both, dealiasing and SVV, to achieve stable (i.e.\ non-crashing) computations. More details on the SVV adopted are reported in section~\ref{subsec:svv}, where we discuss its use to accurately treat under-resolved scales. In this section, we focus on polynomial dealiasing that was a key component without which it was otherwise not possible to obtain stable simulations.

Indeed, the main issue when dealing with element-wise integrations typical of spectral/$hp$ methods, is the inexact numerical integration of the discrete terms of the equations. In fact, it is well known that there is a minimum number of quadrature points $Q_{\min}$, that is required to produce exact results for integrals of non-rational polynomials of any given degree. This minimum number of quadrature points depends on the type of numerical quadrature adopted and on the set of quadrature points used. In our case, we employed Gaussian quadrature on a set of Gauss-Lobatto-Legendre quadrature points~\cite{karniadakis2005spectral}. This type of numerical integration guarantees that the minimum number of quadrature points $Q_{\min}$ necessary to exactly integrate any given order $P$ polynomial $u(\xi) \in \mathcal{P}_P$ up to machine precision is $Q_{\min} = (P + 3)/2$, where $\mathcal{P}$ is the space of polynomials. It is then possible to calculate the number of quadrature points necessary to exactly integrate different nonlinearities that might arise in the equations, as reported in table~\ref{tab:qmin}. We can observe how $Q_{\min}$ increases as the degree of the nonlinearity increases. 
\begin{table}[H]
\centering
\begin{tabular}{c|l}
\toprule
Polynomial order $P$ &  $Q_{\min}$\\
\midrule
$[u(\xi)]^2 \in \mathcal{P}_{2P}$ & $Q  \geq P      + 3/2$ \\
$[u(\xi)]^3 \in \mathcal{P}_{3P}$ & $Q  \geq 3P/2 + 3/2$ \\
$[u(\xi)]^4 \in \mathcal{P}_{4P}$ & $Q  \geq 2P    + 3/2$ \\
\bottomrule
\end{tabular}
\caption{Number of GLL quadrature points for the GLL quadrature to be exact up to machine precision as a function of the polynomial order of the integrand}
\label{tab:qmin}
\end{table}
Let us now consider the incompressible Naver-Stokes equations~\eqref{eq:ns}, and split them into linear and nonlinear terms as follows:
\begin{equation}
\frac{d\mathbf{u}^{\delta}}{dt} = \mathbf{M}^{-1} \left( \mathcal{L} + \mathcal{N} \right) \, ,
\label{eq:discretization}
\end{equation}
where $\mathbf{M}$ is the mass matrix that arises from the numerical discretization, $\mathcal{L} = -\nabla p^{\delta} + \nu \nabla^{2}\mathbf{u}^{\delta} + \mathbf{f}^{\delta}$ contains the linear terms, $\mathcal{N} = -\mathbf{u}^{\delta}\cdot\nabla\mathbf{u}^{\delta}$ contains the convective term, and superscript $\delta$ denotes the spatial discretization. By noting that the convective term is a quadratic nonlinearity, we can identify the $Q_{\min}$ required for exact integration. In particular, given that the spatial discretization we employ is a continuous Galerkin method, we are interested in the integration of $L^2$ inner products of two polynomials $(u_p, u_q)$. On the one hand, to calculate the linear terms, it is necessary to calculate $[u(\xi)]^2 \in \mathcal{P}_{2P}$ (i.e.\ the inner product of the solution with respect to the polynomial basis). Therefore, for exact integration, we need $Q_{\min} = P+2$ Gauss-Lobatto-Legendre quadrature points in one-dimension and a tensor product of points in higher dimensions. On the other hand, to calculate the nonlinear terms, we wish to calculate $[u(\xi)]^3 \in \mathcal{P}_{3P}$, that corresponds to $Q_{\min} = \frac{3}{2}(P+1)$ Gauss-Lobatto-Legendre quadrature points. We note that using other sets of quadrature points, $Q_{\min}$ changes. For instance, the use of Gauss-Legendre points guarantees exact integration up to $Q_{\min} = P + 1/2$, for any given order $P$ polynomial $u(\xi) \in \mathcal{P}_P$. While it is typically possible to exactly integrate the nonlinearities $\mathcal{N}$ present in the equations when dealing with mesh elements that are straight-sided parallelograms, it is more complicated if not impossible when dealing with curved mesh elements. In fact, in the case of straight side parallelograms, the Jacobian of the mapping that defines the coordinates of the element is affine with constant determinant. For curved elements however, this mapping is an isoparametric polynomial expansion, which when incorporated into integrands leads to an additional source of aliasing error, called geometrical aliasing~\cite{mengaldo2015dealiasing}. Geometrical aliasing enters into the discrete formulation via the inverse of the mass matrix $\mathbf{M}^{-1}$, as well as from both the linear and nonlinear terms in equation~\eqref{eq:discretization}, and it is typically in the form of a rational polynomial. It is well known that rational polynomials cannot be exactly integrated within the functional space used to approximate the problem. Therefore, numerical stability cannot be formally guaranteed. We also note that geometrical dealiasing could couple with the underlying equation nonlinearities, thereby exacerbating numerical instabilities. While, for problems that employ unstructured meshes with curvilinear elements like the ones described in this paper, numerical stability cannot be guaranteed, it can be substantially improved by adopting dealiasing strategies that mitigate  aliasing-driven instabilities.

The dealiasing approach adopted in this work is the so-called \textit{Global} dealiasing \cite{mengaldo2015dealiasing}, that employs a larger and identical number of quadrature points on each term of the numerical discretization \eqref{eq:discretization} of \eqref{eq:ns}, that is 
\begin{equation}
\frac{d\mathbf{u}^{\delta}}{dt} = \mathbf{M}^{-1}\Big|_Q \left( \mathcal{L}\Big|_Q + \mathcal{N}\Big|_{Q} \right) \, ,
\label{eq:global-dealiasing}
\end{equation}
The subscript $Q$ in each term of equation~\eqref{eq:global-dealiasing} denotes a higher number of quadrature points than required for exact integration of quadratic linear terms, that is $Q > P + 3/2$. 

This dealiasing strategy mitigates aliasing-driven instabilities induced by equation nonlinearities, curved mesh elements, and the coupling between equation nonlinearities and curved mesh elements. Exact integration is however not guaranteed, except for an asymptotically infinite number of quadrature points. Hence, marginal aliasing errors can still be present and contribute to numerical instabilities. In this work, we always used a quadrature order consistent with the nonlinearities of the governing equations, often with a safety margin to account for curved elements, that is $Q > 3P/2 + 3/2$. We note that for simulations that are adequately resolved (e.g.\ laminar flows), aliasing usually does not affect numerical robustness. However, for under-resolved turbulence computations, this aliasing effect leads to a significant build-up of error in high-frequency solution modes and usually causes the simulation to abruptly diverge.

There are additional tools to suppress aliasing-driven instabilities, that have not been explored in this work. These include the use of skew-symmetric forms of the convective term of the underlying equations~\cite{blaisdell1996effect,gassner2013skew,winters2018comparative}, polynomial filtering~\cite{gottlieb2001spectral,fischer2001filter,fischer2002spectral,hesthaven2008filtering}, variational multiscale schemes~\cite{hughes1998variational,marras2012variational}, and exponential based modal filtering \cite{gassner2013accuracy}. As mentioned above, we note that SVV could, in principle, also be used to suppress aliasing-driven instabilities \cite{karamanos2000aspectral}. However, the diffusion required would be significantly high, and could suppress important flow features, thereby undermining the accuracy of the simulation. Additionally, because the SVV operator is relatively isotropic, whereas dealiasing is anisotropic, the two do not completely overlap. Hence, a reduced amount of regularisation can be achieved with dealiasing, thereby better preserving the solution accuracy.

The use of the dealiasing techniques described in this section removes numerical integration errors that could lead to numerical instabilities in under-resolved simulations; however, it does not address the under-resolved scales that are ubiquitous in high-Reynolds number simulations. These under-resolved scales, similarly to aliasing errors, can be one of the main drivers of numerical instabilities and solution inaccuracies, and they exacerbate aliasing errors leading to the possible failure of the simulation. The handling of under-resolved scales to prevent numerical instabilities and maintain solution accuracy is described next.

\subsection{Regularization for under-resolved scales}\label{subsec:svv}

In order to treat under-resolved scales commonly present in high-Reynolds number simulations, such as the one in section~\ref{sec:car}, we use a regularization technique based on a novel SVV operator. This new operator has been carefully designed to incorporate insights obtained from recent dispersion/diffusion eigenanalysis \cite{moura2016eigensolution,mengaldo2018spatialDG,mengaldo2018spatialFR,moura2020spatial}, and results in well-balanced diffusion characteristics that allow for the accurate simulation of high Reynolds number flow problems. Indeed, when stabilising a high-order formulation, such as the continuous Galerkin scheme adopted in this study, it is critical to achieve an adequate balance between numerical diffusion and solution accuracy, especially for turbulent flow computations.

The general idea behind SVV is that its diffusion can be designed to affect mainly the element-wise polynomial modes of highest order within the spectral/$hp$ element discretization. These polynomial modes comprise the most oscillatory part of the solution and represent the smallest turbulent scales captured by the degrees of freedom employed. The SVV technique was first proposed by Tadmor \cite{tadmor1989convergence} as a stabilisation strategy for classical spectral methods in Fourier space. 
Essentially, if the solution in Fourier space is
\begin{equation}
\textbf{u}(x,t) = \sum_k \hat{\textbf{u}}_k \exp(ikx),
\end{equation}
then SVV ammounts to an artificial diffusion term added to the governing equations in the form
%
\begin{equation}
\nu_{svv} \, \nabla ( \mathcal{Q} \star \nabla \textbf{u} ) = - \nu_{svv} \sum_{\kappa} \kappa^2 \hat{\mathcal{Q}}_\kappa \, \hat{\textbf{u}}_\kappa \exp (i \kappa x) \mbox{ ,}
\label{eq:svv}
\end{equation}

where $\nu_{svv}$ is a constant representing SVV's overall diffusion magnitude, $\kappa$ is the wavenumber, $\hat{\textbf{u}}_\kappa$ are the Fourier coefficients of the solution, $\hat{\mathcal{Q}}_\kappa$ are the Fourier coefficients of the SVV kernel that define how diffusion is to be distributed across Fourier modes, and $i = \sqrt{-1}$. We note that by setting $\hat{\mathcal{Q}}_\kappa \equiv 1$, the standard diffusion (Laplacian) operator is formally recovered. Traditional SVV kernel functions  grow monotonically from zero, achieving $\hat{\mathcal{Q}}_\kappa = 1$ only at the highest order mode. For that solution component, the physical kinematic viscosity $\nu$ effectively becomes $\nu + \nu_{svv}$. 

Since Tadmor's original work on spectral methods, SVV has been adapted to continuous spectral/$hp$ element methods and used in particular to stabilise under-resolved turbulence computations based on the continuous Galerkin method \cite{karamanos2000aspectral,kirby2006stabilisation,koal2012adapting,lombard2016implicit,serson2017direct,}. In this approach the filtering properties of the SVV operator applied to the continuous Galerkin (also referred to as CG hereafter) method are used in-lieu of an explicit turbulence model. The use of techniques originally designed to promote numerical stability in place of a turbulence model is already a common practice, for example, in model-free turbulence simulations via discontinous Galerkin (or DG, hereafter) methods, where upwind-type dissipation is introduced through the Riemann fluxes employed in compressible formulations~\cite{wiart2015implicit,vermeire2016implicit,moura2019implicit}. This setting, where artificial numerical features are used in place of an explicit turbulence model, is commonly referred to as implicit LES, and it has been investigated in the past few decades with various degrees of success~\cite{margolin2002rationale,moura2015modified,chapelier2016study,dairay2017numerical,moura2017setting}. 

The key potential issues when modeling under-resolved scales using the diffusion properties of the underlying numerical discretization or through an SVV-like operator are lack of robustness and/or accuracy of the resulting implicit LES framework, and unclear connections between the artificial numerical filters and the flow physics. In this work, we attempted to address these challenges by encapsulating novel results from the so-called eigensolution analysis (ESA) mentioned previously~\cite{moura2016eigensolution,mengaldo2018spatialDG,mengaldo2018spatialFR,moura2020spatial} to build a new SVV kernel. In this context, we sought for solutions in which the element-wise coefficients arising from the numerical discretization are related to wave-like solutions of the form
\begin{equation}
\mathbf{u} \, \propto \, \exp [i(\kappa x - \omega t)] 
\label{eq:wave}
\end{equation}
through projection (see also \cite{moura2017ontheuse} (Chapter 2).

Initially, we used numerical dispersion and diffusion estimates for spectral/$hp$ element methods in wavenumber/frequency space provided by the so-called  \textit{temporal} ESA~\cite{moura2016eigensolution} to construct an SVV kernel $\mathcal{Q}$ that enforces SVV diffusion to be distributed monotonically across wavenumber space for the CG method. We note in fact that monotonic behavior in wavenumber space is not always guaranteed due the lack of one-to-one correspondence between Fourier and polynomial modes\footnote{Note that common practice was to directly apply Fourier-type kernel entries $\hat{\mathcal{Q}}_\kappa$ upon the orthogonal hierarchical set of element-wise polynomial modes.}. This resulted in what we called ``power-law'' CG-SVV kernel 
\begin{equation}
\tilde{\mathcal{Q}}(p) = \bigg(\frac{p}{P}\bigg)^{P_{svv}} \mbox{ ,} \quad 0 \leq \frac{p}{P} \leq 1 \mbox{ ,}
\label{eq:powerlawkernel}
\end{equation}
where $p$ is the polynomial mode index, $P$ is the overall polynomial order, and $P_{svv}$ is a user-defined parameter that, when increased, further confines CG-SVV damping to the smaller scales. This kernel shape is motivated by hyper-viscosity operators [cite the paper that the reviewer mentions by Avrin and Chang]. We note that the operator $\tilde{\mathcal{Q}}(p)$ in equation~\eqref{eq:powerlawkernel} is formally different from the kernel defined in equation~\eqref{eq:svv}, as this arises from the projection of the element-wise coefficients in equation~\eqref{eq:elemental-expansion} onto the wave-like solution in equation~\eqref{eq:wave}. This kernel also guarantees that raising $P$ increases the resolution power per degree of freedom which is something expected from high-order methods, as discussed in detail in~\cite{moura2016eigensolution}. We highlight that this first study was based on the temporal ESA framework, which is relevant to temporally evolving simulations instead of spatially evolving ones, where inflow and outflow boundary conditions are typically used. While the ``power-law'' kernel defined in equation~\eqref{eq:powerlawkernel} solved a number of issues, still some potentially undesirable characteristics were present. These undesirable aspects were two-fold. On the one hand, non-smooth behavior of the diffusion and dispersion curves was observed for either too small or too large values of $\nu_{svv}$. This could lead to unphysical representation of the flow physics. 
On the other hand, the use of the \textit{temporal} ESA framework did not include spatially evolving flow problems (i.e.\ problem with inflow/outflow boundary conditions) that are commonly found in external aerodynamics. This could trigger spurious reflections of certain waves when they hit regions of mesh expansion (e.g.\ along turbulent wakes), eventually leading to numerical instabilities and divergence of the simulation~\cite{mengaldo2018spatialDG,mengaldo2018spatialFR}.

In order to address the first aspect, we focused on the relevant non-dimensional parameter to quantify SVV's relative intensity is the so-called P\'eclet number, namely $\mbox{Pe} = vh/\nu_{svv}$, that is based on measures of local velocity $v$ and mesh spacing $h$. The adopted strategy was  to maintain the P\'eclet number fixed by making $\nu_{svv} \propto v h$. This led to a CG-SVV scaling that is dimensionally equivalent to the upwind-based dissipation of discontinuous spectral/$hp$ element methods, as explained in~\cite{moura2016eigensolution}. 

To address the second aspect, we pursued a complementary study \cite{moura2020spatial} focused on the so-called \textit{spatial} ESA, which assumes inflow-/outflow-type boundary conditions and is therefore more relevant to spatially developing problems, as the one encountered in this paper. \textit{Spatial} ESA had initially been conducted for discontinuous spectral/$hp$ element methods \cite{mengaldo2018spatialDG,mengaldo2018spatialFR} and proved insightful in explaining how spurious reflections occurred in regions of mesh expansion depending on the Riemann flux employed. In this framework, theory prescribes that an incoming wave of frequency $\omega$ should be associated with a single (and real-valued) wavenumber $\kappa = \omega / a$, $a$ being the advection velocity. However, in a numerical setting, two complex-valued wavenumbers, $\kappa^p$ (physical) and $\kappa^u$ (unphysical), appear in the analysis. Regarding their imaginary part, when numerical diffusion is present, one observes $\Im(\kappa^p) > 0$, which indicates damping of the physical mode during forward propagation, since the solution can be written as 
\begin{equation}
\mathbf{u} \, \propto \, \exp [i(\kappa x - \omega t)] \, = \, \exp [ \, - \Im (\kappa) \, x \, ] \, \exp \{ \, i \, [ \, \Re (\kappa) \, x - \omega t \, ] \} \mbox{ ,}
\end{equation}
where $\Im$ and $\Re$ correspond to the imaginary and real part, respectively. For the unphysical reflected mode, it is instead necessary that $\Im(\kappa^u)$ be negative, indicating damping during backward propagation ($\Delta x < 0$).

\begin{figure}[h!]
\centering 
\includegraphics[keepaspectratio=true, trim={0 0 0 0}, clip, width=0.90\textwidth]{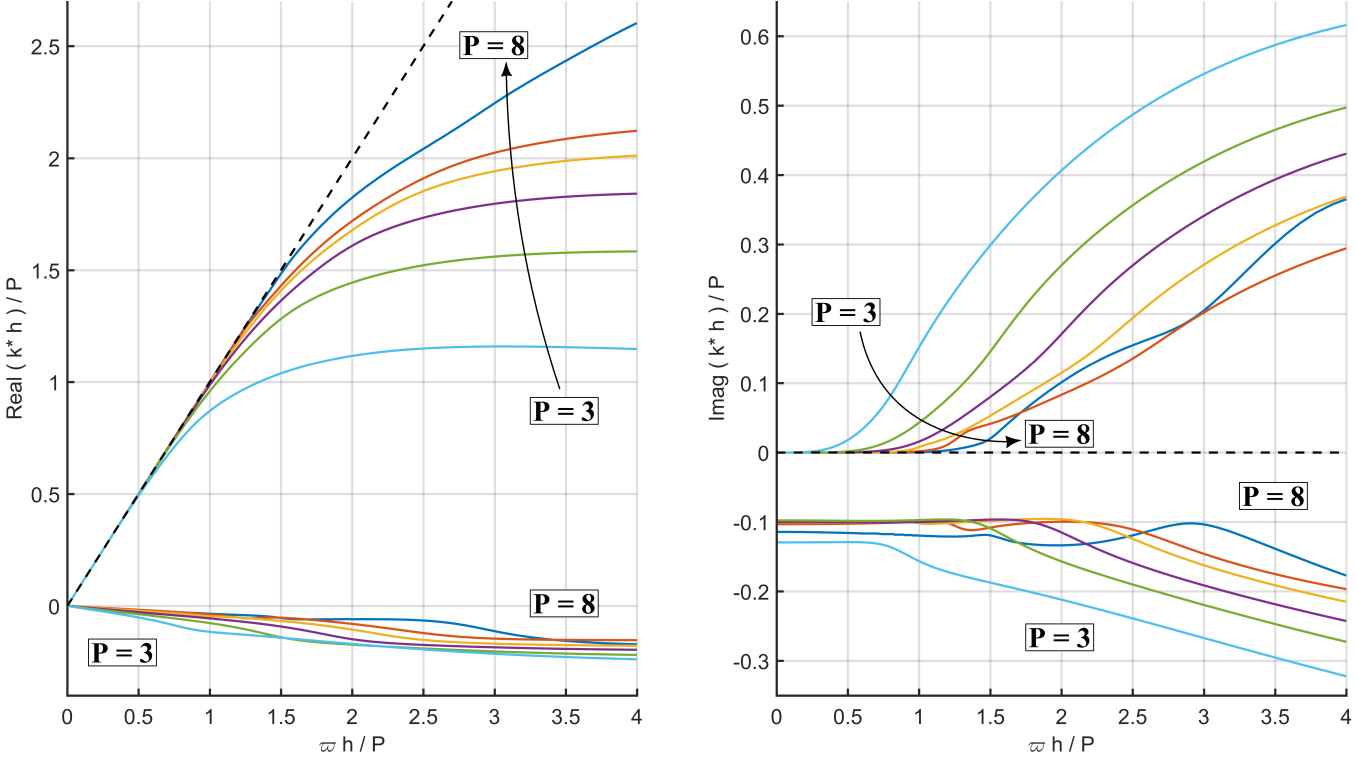}
\caption{Numerical dispersion (left) and diffusion (right) plots for CG with the ``DG kernel'' SVV for various polynomial orders. The unphysical eigenmodes appear on the bottom curves of the graphs. Figure taken from Moura et al.~\cite{moura2020spatial}.}
\label{fig:eigencurves}
\end{figure}

When applied to the CG method in \cite{moura2020spatial}, spatial ESA revealed that unless special care is taken during SVV design, unphysical reflected waves can have negligibly small (although negative) values of $\mathcal{I}(\kappa^u)$. This allows for spurious waves to survive for a long time while interacting with incoming turbulence, eventually causing numerical instability. Hence, a new optimisation procedure was performed for the CG-SVV operator in order to match the diffusion levels of CG to those of standard upwind DG (that uses a Roe-type Riemann solver, in the context of compressible flow problems) at appropriate polynomial orders, this time with a constraint to enforce a sufficiently strong diffusion for the reflected eigenmodes. DG's numerical diffusion was taken as a reference because DG is regarded as having an adequate balance between dissipation and accuracy. The resulting dispersion and diffusion plots are shown in figure~\ref{fig:eigencurves} for various orders, with the curves for the spurious modes at the bottom part of the graphs. Once again, the newly designed kernel for the CG-SVV operator, that we named ``DG'' kernel given its connections to the DG method in terms of diffusion and dispersion characteristics, maintains the P\'eclet number fixed and guarantees that raising $P$ increases the resolution power per degree of freedom.
%

%

\begin{figure}[h!]
\centering 
\includegraphics[keepaspectratio=true, trim={0 0 0 0}, clip, width=0.99\textwidth]{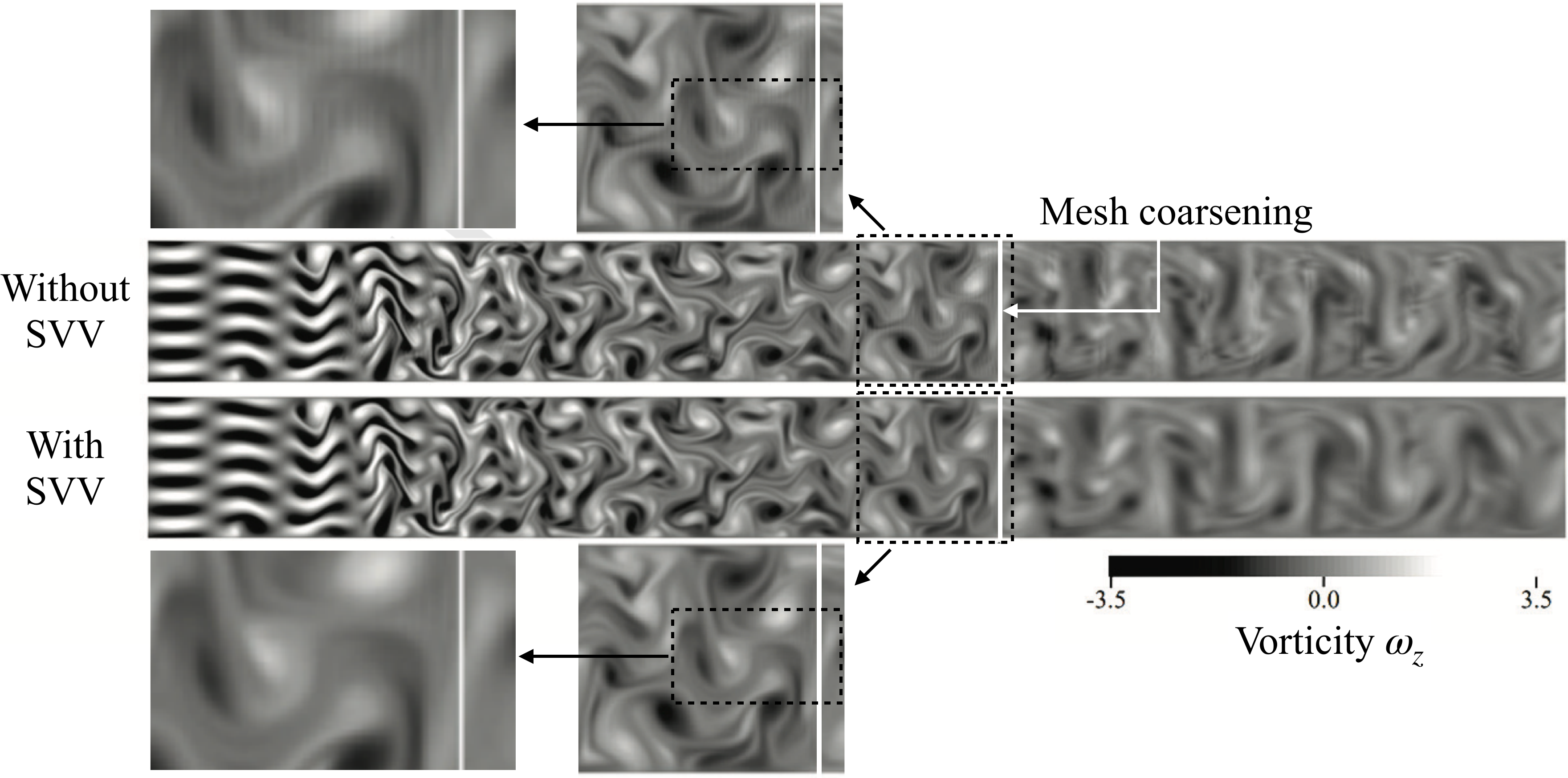}
\caption{Vorticity contours of case $P=8$, $n=11$, at Re = 1000 with (bottom plot) and without (top plot) the novel CG-SVV operator in presence of streamwise mesh coarsening. Close examination shows how the novel CG-SVV kernel is able to suppress spurious reflections originating where the mesh changes resolution. Figure readapted from Moura et al.~\cite{moura2020spatial}.}
\label{fig:gridturb}
\end{figure}

In order to validate the novel CG-SVV operator based on the ``DG'' kernel, we used a simple two-dimensional test case  that was designed to mimic spatially developing grid turbulence \cite{mengaldo2018spatialDG,moura2020spatial}. It consists of a rectangular domain with prescribed unsteady flow at the inlet and symmetry (free-slip) conditions on the sides. This domain was meshed with quadrilaterals in a structured grid pattern featuring two regions. The upstream region had square-shaped elements, whereas the downstream region had a four times larger streamwise mesh spacing. The interface between these two regions marked a potentially unstable threshold zone across which mesh spacing changed abruptly. Results obtained without SVV for $P=8$ and $n=11$ elements in the crossflow direction are shown in figure~\ref{fig:gridturb}. In particular, the vorticity contour plot obtained without SVV, upon close inspection of the upstream region to the mesh coarsening line, shows the presence of spurious reflections originating at the threshold zone. This case diverged shortly afterwards, along with other test cases where we used different polynomial orders, ranging from $P=3$, to $P=7$ (while maintaining the same number of degrees of freedom -- i.e.\ decreasing the number of elements). In particular, at lower polynomial orders, higher Reynolds numbers were required for the simulation to diverge, while for higher polynomial orders, the test cases started to diverge at correspondingly lower Reynolds numbers. When the novel DG-based CG-SVV operator was added, however, up to polynomial order $P=8$, all cases remained stable no matter how high the Reynolds number tested. Furthermore, no spurious reflections have been noticed. And although solutions at lower orders ($P=3$ in particular) might have seemed excessively regularised by the ``DG'' kernel SVV, superior resolution power was effectively recovered at higher orders, such as $P=4$ and above \cite{moura2017eddy}.

The current best practice in $Nektar$ has been to use the ``DG'' kernel in general while employing the ``power-law'' kernel only along homogeneous directions in case periodic boundary conditions are present. In either case, SVV's base coefficient takes the final form
\begin{equation}
\nu_{svv} = (\mbox{Pe}^*)^{-1} \, v \, h / P \mbox{ ,}
\end{equation}
where $\mbox{Pe}^*$ is a reference constant Pecl\'et number (typically of order one), $v$ and $h$ are respectively local measures of advection velocity and mesh size (along the relevant directions of flow propagation), and $P$ is the chosen polynomial order. The entries defining the ``power-law'' kernel are those given in \eqref{eq:powerlawkernel}, whereas those of the ``DG'' kernel have no explicit formula (as the output of an optimisation procedure), having their values given in table form for each order $P$ in \cite{moura2020spatial}. Furthermore, we note that SVV's kernel operation from \eqref{eq:svv} is accommodated in polynomial space as
\begin{equation}
\mathcal{Q} \star \nabla \mathbf{u} \, \approx \, \mathbf{R}^{-1} \;\! \mathbf{Q} \, \mathbf{R} \, \mathbf{\hat{g}} \mbox{ ,}
\end{equation}
in which $\mathbf{\hat{g}}$ denotes the vector of element-wise coefficients in the polynomial expansion of $\nabla \mathbf{u}$, while $\mathbf{R}$ is a ``rotation'' matrix that projects $\mathbf{\hat{g}}$ onto an orthogonal (Legendre) hierarchical basis, and $\mathbf{Q}$ is the diagonal matrix whose entries are given by the chosen SVV kernel function. Therefore, the proposed CG-SVV operator has the final form
\begin{equation}
\nu_{svv} \, \nabla ( \mathcal{Q} \star \nabla \textbf{u} ) \, \approx \, (\mbox{Pe}^*)^{-1} \, \frac{v \, h}{P} \,\, \nabla \left( \, \mathbf{R}^{-1} \;\! \mathbf{Q} \, \mathbf{R} \, \mathbf{\hat{g}} \, \right) \mbox{ .}
\label{eq:cg-svv}
\end{equation}
This CG-SVV operator has been implemented in \nekpp and extensively tested in different test cases of increasing complexity, always displaying strong robustness and preventing numerical instabilities. This approach offers a decent balance between robustness and accuracy for high-order discretizations, being effectively a parameter-free strategy from the practitioners's perspective and allowing for under-resolved turbulence simulations over complex geometries at high Reynolds numbers. This approach, in conjunction with the dealiasing techniques described in section~\ref{subsec:dealiasing}, was used to obtain the numerical simulations depicted in section~\ref{sec:car}. Without the concurrent use of dealiasing and SVV regularization, the numerical simulations presented in this paper would have diverged due to numerical instabilities.

\subsection{Computational efficiency}\label{subsec:comput-efficiency}

One of the advantages of higher order polynomial expansions is their ability to exploit high arithmetic intensities; that is, a greater number of floating point operations is performed for each byte of data retrieved from memory at higher orders than at lower orders. Although this naturally makes operator evaluations at higher order more expensive per degree of freedom, on modern computing hardware this effect frequently tends to be masked, as the cost of transferring data from memory to the compute unit tends to be far greater than that of performing floating point operations. In essence, this can make floating point operations nearly `free' compared to the transfer of data.

However, the efficient implementation of the building-block finite element operators that make up the discretised version of the incompressible Navier-Stokes equations \eqref{eq:ns} that were used in this work, such as inner products or derivatives, still requires careful consideration. Even small changes in polynomial order can yield a large effect on performance due to the many dependent factors, including both the choice of hardware and the problem under consideration. Our recent work in this area~\cite{moxey-2016} has focused on the development of techniques that can automatically determine the most efficient kernels for these operators at runtime. These kernels are aimed to target both clear performance factors (such as operator type and polynomial order), as well as more subtle aspects, such as whether the choice of basis and whether mathematical techniques such as sum-factorisation can be leveraged to increase performance, and the grouping of elements with equivalent numerical properties to improve performance evaluation. The use of these kernels therefore allows an efficient on-node implementation across a broad range of polynomial orders.

For parallelisation between nodes, \nekpp uses a parallelisation model based on distributed MPI. The METIS library~\cite{karypis1998fast} is first used to partition the domain based on the dual graph representation of the mesh. Since, for industrial geometries, the mesh generation process yields a hybrid mesh of prismatic and tetrahedral elements, these are weighted appropriately in the partitioning process, so as to approximately equally distribute the work between processors. Both the pressure Poisson equation and the three scalar velocity correction Helmholtz equations that comprise the operator splitting scheme are solved implicitly, leading to the need for efficient solvers for large, distributed linear algebraic positive definite systems. To achieve this, we apply an iterative preconditioned conjugate gradient (PCG) method for each scalar system. We note, however, that the usual approach taken at lower orders of representing this system as a large, distributed sparse matrix is not generally efficient at higher orders. Instead, when the action of the matrix multiplication is required during PCG iterations, we implement this as many smaller, locally dense elemental matrices, combined with a sparse gather operation to apply $C^0$ connectivity. The gather operation uses the efficient {\tt Gslib} library in parallel, based on the code Nek5000~\cite{fischer2008nek5000}. Our tests have demonstrated that this combination allows both performance and scalability of the PCG solver up to $10^5$ cores at high levels of efficiency. Finally, we discuss preconditioning of these methods. All four PCG solvers use an effective and scalable low-energy basis preconditioner \cite{sherwin2001low,grinberg2009parallel} and, optionally, projected initial conditions based on storage from solutions at previous timesteps~\cite{fischer1998projection}. The solution of the pressure Poisson equation, however, is particularly ill-conditioned, and therefore requires the use of coarse-space preconditioning, which requires the inversion of the linear finite element representation of the mesh. Currently, the coarse space preconditioner uses a direct parallel inverse method developed by Fischer and Tufo~\cite{tufo2001}.

The combination of all of these intra- and inter-node performance optimisations, together with a robust meshing pipeline and stabilisation strategies, has yielded an efficient solver at high orders that is capable of tackling complex configurations required for industrially relevant problems at superior computational performance.

%
\section{High-order iLES for the aerodynamic (re)design of the Elemental Rp1 road car}
\label{sec:car}
%

\begin{figure}[htbp!]
\begin{center}
\begin{tabular}{ccc}
\includegraphics[width=0.45\textwidth]{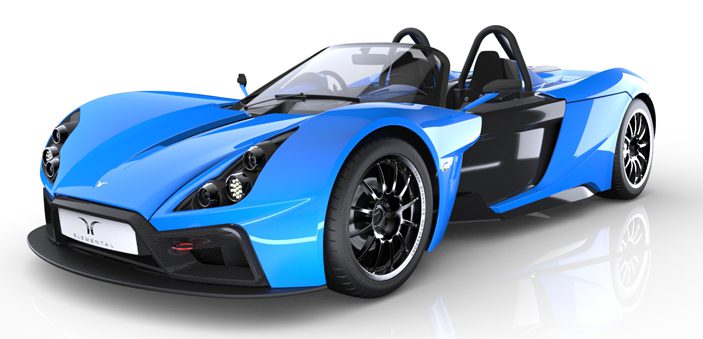} && \includegraphics[width=0.40\textwidth]{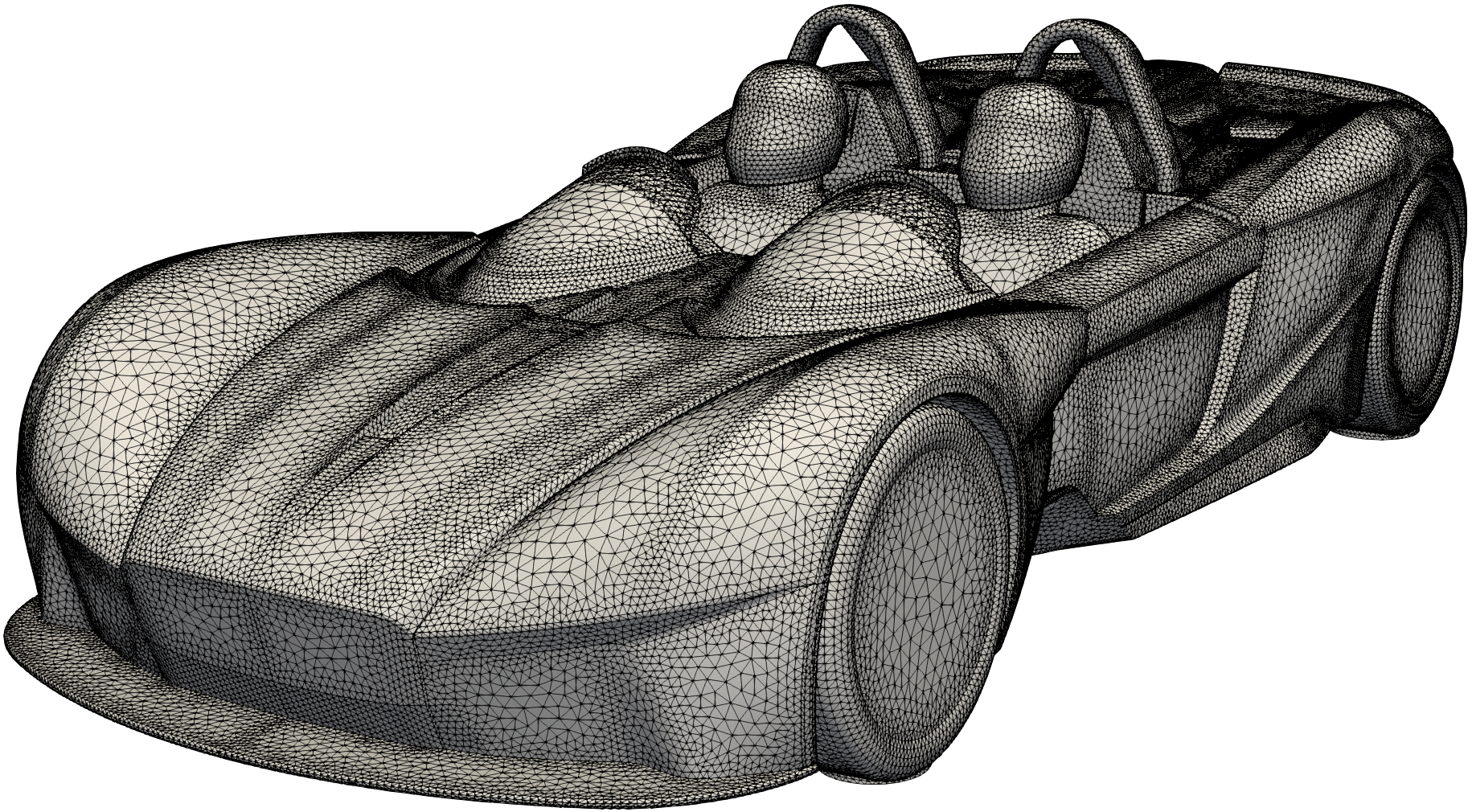} \\
(a) && (b)
\end{tabular}
\end{center}
\caption{The original design of the Elemental Rp1 road car, a high performance, street legal track car which delivered 400~kg of downforce at 150~mph.}
\label{fig:rp1-car}
\end{figure}

The developments in terms of high-order mesh generation and numerics implemented in \nm and \nekpp that were described in sections~\ref{sec:mesh} and \ref{sec:num-met} were used to perform large-scale high-fidelity LES simulations of a real road car, namely the Rp1 car~\cite{RP1}. The Rp1 car, produced by the British manufacturer Elemental Cars since June 2016, is a high-performance, light-weight, road-legal vehicle that weighs 580~kg. It is equipped with a 320~bhp engine and accelerates from 0 to 60~mph in just 2.6 seconds. Its current configuration, depicted in Figure~\ref{fig:rp1-car} along with its digital twin, delivers 400kg of downforce at 150 mph. The main targets of the high-fidelity LES simulations performed were to analyze and improve the external aerodynamic performance of the car. In particular, we analyzed three configurations, namely designs 1, 2, and 3, that we denote as D1, D2, and D3, respectively. D1 is the baseline design, while D2 and D3 are two design upgrades that were implemented to improve aerodynamic performance and pilot comfort. This study was a collaborative effort put forward by Imperial College London, Elemental cars group~\cite{elemental}, London Computational Solutions (LCS)~\cite{lcs} and Silicon Graphics International Corp (now Hewlett Packard Enterprise). All the simulations were carried out using the open-source library \nekpp. In the following, we describe the configuration of the simulations and we discuss qualitatively the results obtained. 

\subsection{High-order meshing strategy and simulation configuration}
This section outlines the details of the meshes adopted along with the strategies used to generate them (section~\ref{subsec:rp1-mesh}), and discusses the setup of the simulations performed (section~\ref{subsec:rp1-setup}).

\subsubsection{Mesh generation}\label{subsec:rp1-mesh}
The hugely complicated nature of the geometry under consideration, which involves hundreds of CAD surfaces and thousands of edges, presented a significant test for the mesh generation process outlined in section~\ref{sec:mesh}. The high-order meshes used in this study ranged from approximately 2.2 million  to 4 million elements, and they are comprised of a combination of triangular prisms in the near-wall boundary layer region with 6 layers used to resolve separation accurately, and tetrahedra in the rest of the domain. 

We adopted three different meshing strategies for the three car designs, D1, D2, and D3. In particular, we initially used the first and second \textit{analytic curving} method described in section~\ref{subsec:linear-mesh}, for the surface mesh generation stage. The first \textit{analytic curving} method was applied to the first design, namely D1, where the creation of a viable mesh took approximately 4 months of iterations between CAD healing and high-order meshing. The final mesh was composed of 2.2 million elements, including 600000 prisms within the boundary-layer mesh. The second \textit{analytic curving} method was instead applied to the second design, namely D2, that was constructed based on the results from D1. The mesh for D2 contained 2.6 million elements, including 700000 boundary layer prisms. While, both \textit{analytic curving} strategies were effective in generating viable meshes for D1 and D2, they were deemed excessively user-driven and time-consuming. The mesh for D3 was obtained with what we named the \textit{projection curving} method, described in section~\ref{subsec:linear-mesh}. Indeed, this method, along with the use of \cf for geometry repair, \ccm for linear and boundary-layer mesh generation and \nm for high-order mesh generation was the least time-consuming and user-driven. On the one hand, it allowed cutting the overall meshing time from 4 months to 1 week. On the other hand, the mesh was created almost without user effort, and the process required one single execution of linear mesh generation via \ccm and one single execution of \nm. No repetitive cycles of CAD healing were required. The final mesh produced for D3 contained approximately 3.5 million elements, including 1.24 million boundary-layer prisms. Although the meshing strategy for D3 is the most robust and efficient, we should point out that allows for inaccuracies in the geometry description, as a small percentage of surface elements are included in the ``exclusion set'', thereby remaining straight-sided. However, in the results that we present in section~\ref{subsec:second-design}, no obvious unphysical results could be noticed, and the flow physics in those regions of the car that were geometrically identical between D2 and D3 was indistinguishable.

\begin{figure}[htbp!]
\begin{center}
\begin{tabular}{ccc}
\includegraphics[width=0.35\textwidth]{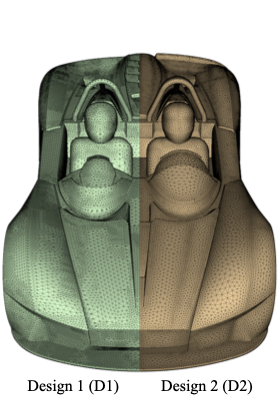}  & \hspace{1cm} & \includegraphics[width=0.35\textwidth]{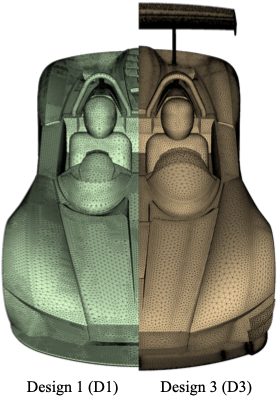} \\
(a) & \hspace{2cm} & (b)
\end{tabular}
\end{center}
\caption{Surface meshes of the three car designs considered, D1, D2, and D3. Figure (a) shows a comparison between D1 and D2, while figure (b) shows a comparison between D1 and D3. It is possible to notice a smoother distribution of surface elements for both D2 and D3 when compared to D1. This is due to the different approaches adopted for the surface mesh generation, where for D1 we used the first \textit{analytic curving}, while for D2 and D3 we used the second \textit{analytic curving} and \textit{projection curving}, respectively.}
\label{fig:meshes_d1_d2_d3}
\end{figure}

Figure~\ref{fig:meshes_d1_d2_d3} shows the three surface meshes for D1, D2, and D3 obtained via the three different strategies just outlined. It is possible to note that the surface mesh for D2 and D3 has a smoother element distribution than D1, due to the enhanced triangulation strategies adopted, namely the second \textit{analytic curving}, and \textit{projection curving}. We remark that for all the simulations carried out we did not include the spokes that are instead shown in figure~\ref{fig:rp1-spokes}.

\subsubsection{Simulations setup}\label{subsec:rp1-setup}

The simulation setup required to analyze the aerodynamic of the Rp1 car was challenging due to the high Reynolds number, $Re=10^{6}$, based on the length of the car $L$, and to the complex geometries involved. In order to obtain stable yet accurate simulations, we adopted some simplifications that are commonly adopted when performing the aerodynamic car simulations. 

The Rp1's unique double diffuser design and the the road-tyre interaction were carefully modeled with some simplifications that allowed a significantly easier simulation setting while not undermining the accuracy of the results. Rp1's double diffuser design has intakes and outlets on the body of the car. To model these accurately, we extracted the corresponding surfaces and constructed an appropriate velocity boundary field, by enforcing a prescribed average normal velocity on each surface. The boundary condition has to also satisfy the no-slip condition at the edges of the inlet surfaces and this was imposed by solving a Helmholtz problem where the positive Helmholtz coefficient was chosen to provide a layer of desired thickness at  the edges. This was possible by using a manifold solver developed by Cantwell et al.~\cite{cantwell2014}. In addition, the road-tyre interaction was modeled using a contact patch along with rotational boundary conditions to a simplified tyre geometry. This allowed efficiently modeling the rotation of the tyre without the need for a moving mesh. Finally, we simulated only half of the car, by using a symmetry plane. These simplifications allowed a faster simulation setup and simulation runtime, without altering the flow features that were of interest in this study. 

The simulations were initialized adopting a similar strategy to Lombard et al.~\cite{lombard2016implicit}, where we began at a lower Reynolds number, $Re=10^{4}$. The initial conditions were imposed impulsively so that $u/U=1$, where $U$ is a reference velocity, and all non-moving surfaces are treated as no-slip boundaries. At the outflow, we imposed the boundary condition developed by Dong et al.~\cite{dong2014}, which balances the kinetic energy influx through the outflow boundary condition to prevent instability. The Reynolds number is then gradually increased by a factor of ten until the target Reynolds number, $Re = 10^{6}$, is reached. Between increases, the flow is evolved for two convective time units $t_{c} = L / U$ to allow the flow to adapt to the new flow physics dictated by the higher Reynolds number. We note that the Reynolds number used in this study, $Re = 10^{6}$, is still lower than the experimental value of approximately $Re = 2\times10^{6}$. Nevertheless, to the authors' knowledge, this is the highest Reynolds number ever attempted for a real car geometry, using iLES in conjunction with spectral/$hp$ element methods. 

The simulations were run at a polynomial order 4 (fifth order of accuracy) with a second order implicit-explicit time-stepping. To increase the accuracy of the integration, a seventh order of quadrature was used, that consisted of the so-called global consistent integration introduced in Mengaldo et al.~\cite{mengaldo2015dealiasing}, and described in section~\ref{subsec:dealiasing}. This allowed us to mitigate aliasing-driven instabilities. We note that the use of a higher quadrature meant that the meshes were generated at a higher order of $P=7$ to ensure the elements were valid with this quadrature. In addition, the SVV kernel presented in section~\ref{subsec:svv} was used to filter the under-resolved scales and enhance the stability of the simulations. Without the use of both, consistent integration and SVV was not possible to obtain numerically stable calculations.

In terms of computational requirements, the simulation were performed in parallel using 1000 to 4000 cores, depending on node availability on a SGI UV machine, and it took approximately 3 days to evolve the flow for one convective length. The high-fidelity flow simulations performed allowed unique insights into the aerodynamic performance of the Rp1's complex geometry, highlighting a number of important regions for aerodynamic optimisation. These results are discussed next.

\subsection{Results}\label{subsec:results}
In this section, we discuss the results obtained on the three car designs introduced earlier, D1, D2 and D3. In particular, in section~\ref{subsec:first-design}, we present a comparison between D1 and D2, while in section~\ref{subsec:second-design}, we present a comparison between D1 and D3. The results focus on iso-contour quantities that allowed us to identify important flow features suitable to optimization.

\subsubsection{First design stage}\label{subsec:first-design}

\begin{figure}[htbp!]
\begin{center}
\begin{tabular}{ccc}
\includegraphics[width=0.35\textwidth]{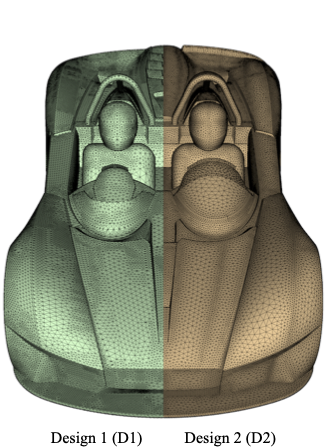}  & \hspace{2cm} & \includegraphics[width=0.35\textwidth]{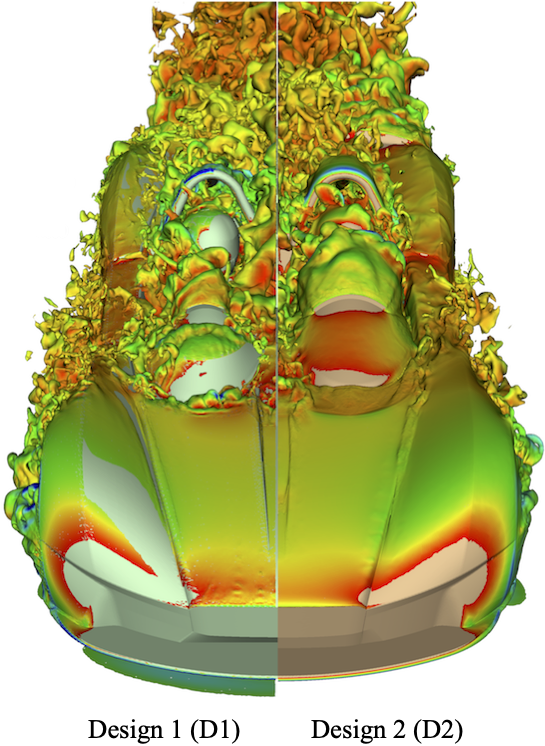} \\
(a) & \hspace{2cm} & (b)
\end{tabular} \\[1em]
\includegraphics[width=0.70\textwidth]{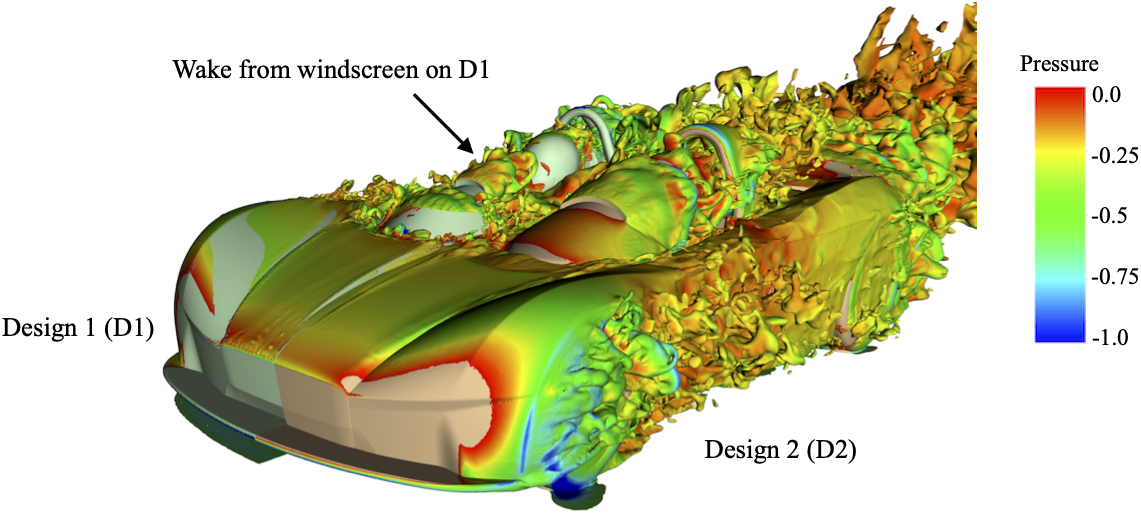} \\
(c)
\end{center}
\caption{Comparison between D1 and D2, in terms of mesh [figure (a)], and flow physics [figures-(b) and -(c)]. The images depicting the flow physics shows the total pressure coefficient for $Cp_{0}=0$ and the iso-contours are colored in pressure.}
\label{fig:d1_vs_d2}
\end{figure}

The first simulation performed was for design 1 (D1), that is the first baseline model of the car that was produced, followed by the simulation of design 2 (D2). The latter design, D2, implements an upgrade of the aerodynamic package with respect to D1. Figure~\ref{fig:d1_vs_d2} shows the meshes and associated results for D1 and D2. In particular, figure~\ref{fig:d1_vs_d2}-(a), shows the meshes for the two designs (that are the same as depicted in figure~\ref{fig:meshes_d1_d2_d3}), while figure~\ref{fig:d1_vs_d2}-(b) shows the associated flow results. Figure~\ref{fig:d1_vs_d2}-(c), shows the same flow results as the right image in figure~\ref{fig:d1_vs_d2}-(a), from a sideview perspective. The flow results depicted in figure~\ref{fig:d1_vs_d2} show iso-contours of the total pressure coefficient $Cp_{0} = 1 - (u / u_{\infty})^2 = 0$, and they are colored as pressure.
From the high-order D1 simulation two key findings were made, neither of which were identified in low-order RANS simulations \cite{Taylor2017}. Firstly, there appeared to be strong vortical structures hitting the drivers helmet (see figures~\ref{fig:d1_vs_d2}-(b) and \ref{fig:d1_vs_d2}-(c) for D1). Indeed, aerodynamic disturbances were also reported by the test drivers, confirming the finding of our iLES. These vortical structures were generated by the windscreen, and the issue was fixed in D2 with a redesigned console area in front of the driver, significantly enhanced the driving experience. The D2 iso-contour shows clearly that these structures now pass cleanly over the drivers head and are less noisy in terms of small-scale vortices (figures~\ref{fig:d1_vs_d2}-(b) and \ref{fig:d1_vs_d2}-(c), for D2). Secondly, the roll hoop produced significantly more drag and separation than predicted. D2 had a redesigned roll hoop with an aerofoil profile as opposed to a circular cylinder. This aerofoil profile was slightly angled to help control the flow over the new Gurney flap at the rear of the car. This, combined with redesigned diffusers on the underside, led to increased downforce over D1. The trend of changes in downforce and drag between the D1 and D2 simulations was well predicted by the high-order simulations and matched the trend in the RANS results \cite{Taylor2017}.

\subsubsection{Second design stage}\label{subsec:second-design}

The final simulation performed in this study is for design 3 (D3), which is a full aerodynamic upgrade over the previous two designs, D1 and D2. This car is designed to achieve extremely high levels of downforce specifically for track racing. This design includes a fully redesigned floor, front splitter and the addition of a rear wing. The ride height has also been altered, raising the car at the rear for increased diffuser performance and lowering the front of the car to increase in-ground effect of the front splitter. 

\begin{figure}[h]
\begin{center}
\begin{tabular}{ccc}
\includegraphics[width=0.31\textwidth]{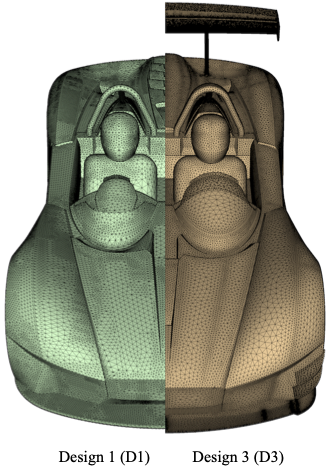}  & \hspace{2cm} & \includegraphics[width=0.31\textwidth]{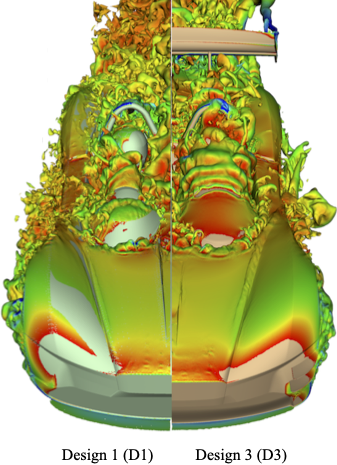} \\
(a) & \hspace{2cm} & (b)
\end{tabular} \\[1em]
\includegraphics[width=0.6\textwidth]{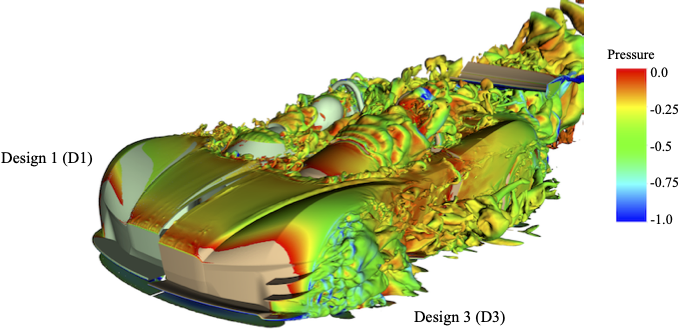} \\
(c)
\end{center}
\caption{Comparison between D1 and D3, in terms of mesh [figure (a)], and flow physics [figures (b) and (c)]. The images depicting the flow physics shows the total pressure coefficient for $Cp_{0}=0$ and the iso-contours are colored in pressure.}
\label{fig:d1_vs_d3}
\end{figure}

Figure~\ref{fig:d1_vs_d3}, similarly to figure~\ref{fig:d1_vs_d2}, shows again the total pressure coefficient, colored by pressure.
The results are consistent with the D2 simulation and are shown alongside the D1 results. The trends in increasing downforce were well predicted by the high-order simulation. In figure~\ref{fig:d1_vs_d3}, we note that it is possible to observe a noticeable offset in the geometries of the two cars, due to the alteration in ride height.  

Although we used an approach for meshing D3 that can be aggressive in preserving mesh validity versus geometrical accuracy, thereby leaving a non-negligible number of elements straight-sided as opposed to curvilinear (the number of straight-side elements was $<5\%$), the flow physics is in agreement with D1 and D2, and no unphysical results could be readily noticed. However a more comprehensive study should be undertaken in order to draw stronger conclusions on whether the compromised geometric accuracy of our most robust meshing pipeline is a compromise worth taking. The early results here show that it may well be.

%
\section{Conclusions}\label{sec:conclusions}
%

This work demonstrates the significant progress in leveraging \nekpp for the goal of moving high-order CFD from academia to industry, and some of the benefits that high-order methods can provide to the industrial community. Significant developments had to be made in order to overcome obstacles in both mesh generation and solver technologies. 

The novel high-order meshing workflow described in section~\ref{sec:mesh} allowed the robust, efficient and accurate high-order mesh generation of an extremely complex geometry. This was a key to prevent unphysical diffusion being generated at no-slip surfaces that constitute the geometry of the car, and to avoid using an excessive number of linear elements in proximity of curved geometrical features. Indeed, the meshing framework has proven reliable and viable for industrial geometries, and it was a key enabler to successfully accomplish the simulations presented in section~\ref{sec:car}.
 
Along with the high-order meshing framework, the novel numerical technologies described in this section~\ref{sec:num-met} were essential to successfully produce the CFD workflow described in this paper. More specifically, the consistent integration for mitigating aliasing-driven instabilities in conjunction with the CG-SVV based on the ``DG'' kernel for treating under-resolved scales were the keys to obtain stable simulations and accurate simulations. The use of the novel CG-SVV approach, in particular, was instrumental to produce high-fidelity results. In fact, the CG-SVV operator is based on ESA's estimates for spectral/$hp$ element methods, where numerical dispersion and diffusion characteristics are quantified in wavenumber/frequency space, as in classical von Neumann analyses. Although mostly feasible for simple model problems, such as one-dimensional linear advection, ESA's estimates for spectral/$hp$ methods have been shown to hold more generally for nonlinear problems and even turbulence simulations \cite{moura2015linear,moura2017eddy,fernandez2019nonmodal}. These estimates quantify how the numerical error (diffusion in particular) increases at large wavenumbers/frequencies, affecting especially the smallest, poorly- and under-resolved scales. When interpreted alongside implicit LES experiments, ESA reveals that spectral/$hp$ discretisations of higher polynomial orders perform best because they affect the large and intermediate flow scales as little as possible, while gradually introducing a sufficiently strong diffusion to regularise the smaller scales. This way, the governing equations and thus the turbulence physics are solved more accurately for a wider range of scales, rather than having to rely on (often restrictive) modelling hypothesis. We note that past studies also focussed on how numerical diffusion acts as a turbulence model~\cite{margolin2002rationale,moura2015modified,chapelier2016study,dairay2017numerical,moura2017setting}. However the ESA framework was deemed the most promising route to achieve high-fidelity implicit LES results.

The simulations in section~\ref{sec:car} represent the first high-order (having solution polynomials $= 4$) accurate results of a real automotive geometry by means of implicit LES technologies using spectral/$hp$ element methods. They represent one of the first attempts at bridging the gap between applied mathematics academic research and industrial applications within the context of CFD.

\section*{Acknowledgments}
This work has benefited from the technical support and design insight from Elemental Cars and from access to large HPC resources provided by Hewlett Packard Enterprise. We also would like to thank Mark Gammon and his group at ITI Global for providing us with a license to the program CADfix and its CFI interface, and for their technical help and support. Partial financial support was provided by EPSRC under the Platform Grant PRISM: Platform for Research In Simulation Methods (EP/R029423/1) and by the EU Horizon 2020 project ExaFLOW (grant 671571). M.~Turner acknowledges the support of EPSRC and Airbus under an Industrial CASE studentship. We also acknowledge support from Imperial College High Performance Computing Service.

\bibliographystyle{siamplain}
\bibliography{references}
\end{document}